\documentclass[a4paper,fleqn]{cas-sc}
 \usepackage[numbers]{natbib}

\usepackage{multirow}
\usepackage{subfig}
\usepackage{float}
\usepackage{adjustbox}
\usepackage{rotating}
\usepackage{lscape}
\usepackage{float}
\usepackage{graphicx}

\usepackage{makecell}
\usepackage{comment}
\usepackage{caption}
\def\tsc#1{\csdef{#1}{\textsc{\lowercase{#1}}\xspace}}
\tsc{WGM}
\tsc{QE}
\tsc{EP}
\tsc{PMS}
\tsc{BEC}
\tsc{DE}

\begin{document}
\let\WriteBookmarks\relax
\def\floatpagepagefraction{1}
\def\textpagefraction{.001}

\shorttitle{ }


\title[mode = title]{DBE-KT22: A Knowledge Tracing Dataset Based on Database Systems Exercises}                      



%
\author{Ghodai Abdelrahman (1)*, Sherif Abdelfattah (2), Qing Wang (1), Yu Lin (1) \\(1) School of Computing, Australian National University, Australia\\(2) School of Engineering and Information Technology, University of New South Wales, Australia }[
                        orcid=0000-0002-7470-0494]



\ead{ghodai.abdelrahman@anu.edu.au}

\begin{abstract}
Online education has gained an increasing importance over the last decade for providing affordable high quality education to students worldwide. This has been further magnified during the global pandemic as more students switched to studying online. The majority of online education tasks, e.g., course recommendation, exercise recommendation, or automated evaluation, depends on tracking students' knowledge progress. This is known as the \emph{Knowledge Tracing} problem in the literature. Addressing this problem requires collecting student evaluation data that can reflect their knowledge evolution over time. In this paper, we propose a new knowledge tracing dataset named  \emph{Database Exercises for Knowledge Tracing} (DBE-KT22) that is collected from an online student exercise system in a course taught at the Australian National University in Australia. We discuss the characteristics of the DBE-KT22 dataset and contrast it with the existing datasets in the knowledge tracing literature. Our dataset is available for public access through the Australian Data Archive platform upon proper citation to the authors.\footnote{\url{https://dataverse.ada.edu.au/dataset.xhtml?persistentId=doi:10.26193/6DZWOH}}
\end{abstract}




\ExplSyntaxOn
\keys_set:nn { stm / mktitle } { nologo }
\ExplSyntaxOff
\maketitle

\section{Introduction}
\label{Sec:IN}

Over the last decade, online education systems rivaled conventional classroom-based education for the ease of accessibility worldwide using the Internet, high quality of learning materials, and affordable cost. The recent global pandemic further amplified the impact of online education as an effective alternative that could overcome physical distancing restrictions imposed on students and teaching staff in schools and university campuses. Nevertheless, one of the significant challenges that need to be addressed in online education systems is the ability to effectively trace a student's learning progress, similar to what a human teacher would do in the classroom. Human teachers rely on their intuition and experience to estimate a student's knowledge state and tailor the learning process accordingly. 

Acquiring such ability would enable online education systems to archive many vital education objectives, including customized curriculum generation, learning materials recommendation, exercise recommendation, automatic evaluation, or learning feedback generation. Achieving such objectives would facilitate automating the teaching process and pave the way for transforming the current online education systems into \emph{Intelligent Tutoring Systems} (ITS). An ITS not only automates the teaching procedure using computer systems (e.g., web applications) but also handles supporting tasks such as customizing the learning experience and providing guidance and feedback to the students \cite{ITS_review21}. 

\begin{figure}
\centering
\includegraphics[width=0.9\textwidth]{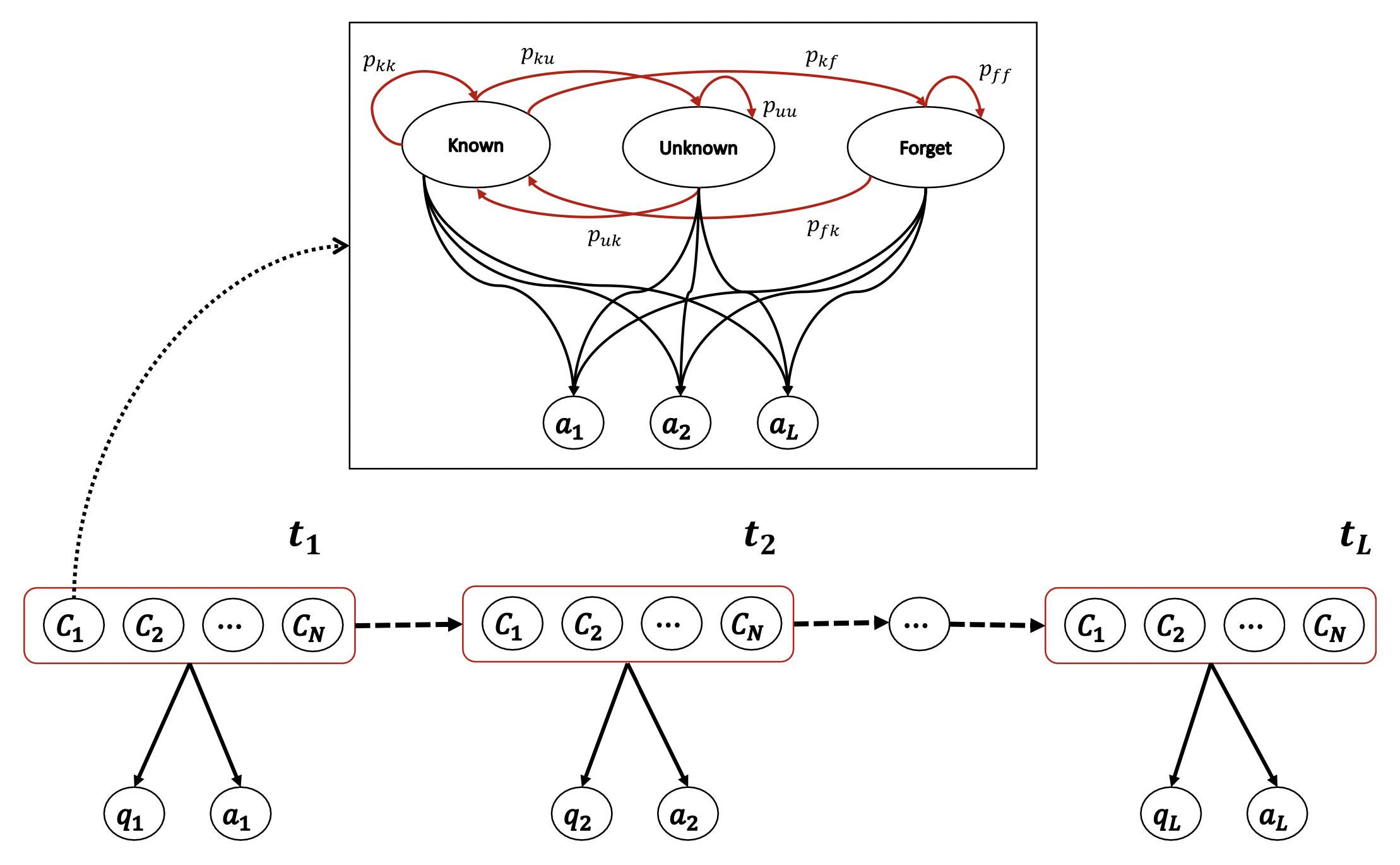}
\caption{A probabilistic graphical model representing a knowledge tracing scenario.}
\label{fig:kt_gmodel}
\end{figure}

The \emph{Knowledge Tracing} (KT) problem formulates the challenge of tracing a student's knowledge state based on their exercise answering history \cite{piech2015deep, kt_survey22}. In particular, the exercise answering history could be represented as a sequence of question-answer pairs, and the task of a solving computational model would be to predict the likelihood of correctly answering the following questions. Figure~\ref{fig:kt_gmodel} depicts a probabilistic graphical model for a KT scenario. At each time point $i$ in a previous answering sequence that spans from time $t_1$ to $t_L$, we observe a question tag $q_i$ (i.e., question text) and an answer $a_i\in\{0,1\}$. A student knowledge state at a given time point is represented by a vector over the proficiency states of the involved learning concepts $[C_1,C_2,\dots,C_N]$, where a learning concept $C_n$ could be a topic in a course such as a number addition, or subtraction in an elementary math subject. It has to be noted that the KT literature tends to refer to learning concepts as knowledge components (KCs)~\cite{kt_survey22} as they constitute the components of a student's knowledge state. At the top part of Figure~\ref{fig:kt_gmodel}, we show a hidden Markov model representing the modes that an individual learning concept's proficiency state could take through the learning process, including \emph{known} for a sufficient knowledge, \emph{unknown} for insufficient knowledge, and \emph{forget} for forgetting what has been known before on this concept.

Figure~\ref{fig:kt_gmodel} shows that the KT problem is a challenging one that involves a hierarchical dynamical system that consists of two levels, including the individual proficiency states of KCs and the overall student's knowledge state. Early attempts to address the KT problem date back to three decades ago, where Bayesian inference methods were used to predict the likelihood of correctly answering a new question given the previous answering sequence~\cite{corbett1994knowledge}. Recently, deep learning approaches~\cite{piech2015deep,Going16} have shown a promising potential in addressing the KT problem due to the representation capacity of deep neural networks and the use of advanced sequence modeling techniques such as differential associative memory~\cite{zhang2017dynamic, SKVMN} to capture long-term dynamics in the exercise answering sequence, graph neural networks~\cite{TongLHHCLM020, GIKT20, GKT19, abdelrahman2021deep} for incorporating relationships among learning concepts and questions, and attention~\cite{pandey2020rkt, SAKT19, AKT20, SAINT20} for focusing on relevant questions in the previous answering sequence. The performance of deep neural models is significantly impacted by the volume and quality of available training data; thus, there is a need for datasets that could reflect the realistic dynamics in the online learning processes.

Despite of the availability of relevant KT datasets~\cite{feng2009addressing,STATICS2011,Junyi,piech2015deep,EDNET2020ednet,kdd2010}, there are limitations that burden the use of these datasets in evaluating advanced deep KT models including (1) missing text data for question tags and description text for the involved learning concepts, which limits the use of advances in deep language models that could find an effective question embedding representation, (2) missing ground-truth on the relationships among learning concepts, which could be used by deep graph neural networks to incorporate the influence between learning concepts when updating the knowledge state, (3) missing ground-truth on question difficulty, which could provide an auxiliary signal when predicting the likelihood of correct answer, (4) lack of data for advanced student groups such as undergraduate and graduate students as the majority of datasets were collected from school-grade students, (5) lack of insights on the student's uncertainty while answering a given question, which could provide features to quantify the likelihood of important answer aspects such as forgetting, guessing, or slip (i.e., mistaken answer).

To address these limitations, we propose a new dataset named \emph{DBE-KT22}, which was collected from an introduction to relational databases taught at the Australian National University (ANU) in Australia for undergraduate and graduate students across multiple disciplines, including computer science, engineering, arts, and business between 2018-2021 academic years. The contributions of the DBE-KT22 dataset can be summarized as follows:

\begin{itemize}
    \item Providing detailed meta-data for questions, including tag text, hints text, choice text, and images that provide multi-model feature sources for question embedding representation learning.
    \item Providing detailed meta-data for the involved learning concepts, including learning concept tag text and description text to facilitate learning representative learning concept embeddings.
    \item Incorporating the ground truth on the relationships between learning concepts and each other and between learning concepts and questions in the form of graphs to facilitate the use of graph representation learning methods.
    \item Presenting the ground truth on question difficulty provided by domain experts teaching the subject over a long time period. This provides auxiliary features that could be used for answer prediction and evaluating methods targeting quantifying the question difficulty, such as question recommendation and curriculum generation models.
    \item Capturing different aspects that reflect a student's uncertainty while answering questions, including self-feedback on question difficulty, self-feedback on confidence in the answer, indication for using hints, number of times a student changes their answer, and the total time taken to answer. This dataset enables answer uncertainty features in addition to the features from the previous question-answering sequence. 
    
\end{itemize}

The remainder of this paper is organized as follows. Section~\ref{Sec:DA} provides a detailed review of the available KT datasets and their characteristics. Section~\ref{sec:anukt22} introduces the details of the proposed DBE-KT22 dataset, including data collection, distribution, and formatting. Section~\ref{sec:experiments} presents our experimental analysis to highlight the characteristics of the proposed dataset. Finally, Section~\ref{sec:conclusion} concludes the work.
\section{Related Work}
\label{Sec:DA}
To facilitate the use of machine learning and statistical models in addressing the KT problem, many specialized datasets have been proposed in the literature. The early attempts to generate and publish a KT dataset date back three decades ago~\cite{corbett1994knowledge} with the rise of statistical KT models. Since then, the collected data's characteristics and volume have been enriched to cope with the need for large training datasets demanded by deep learning models. This section introduces the characteristics of well-known KT datasets in the literature and highlights their differences. Table~\ref{tbl:CD} summarizes each dataset's key statistical characteristics and meta-data, including our proposed one. We detail each of the comparative datasets as follows. 

\subsection{ASSISTments Datasets}
This is one of the most popular groups of datasets in the KT literature. The ASSISTments datasets~\footnote{\url{https://sites.google.com/site/assistmentsdata}}~\cite{feng2009addressing,pardos2014affective} were collected in the time period $2009-2015$ using an online education platform with the same name\footnote{\url{https://www.assistments.org/}}. In particular, the data was recorded based on a high school Math assessment sampled from the \emph{Massachusetts Comprehensive Assessment System} (MCAS)\footnote{\url{https://www.doe.mass.edu/mcas/testitems.html}}. We introduce each variant of this group of datasets below. 

\begin{itemize}
    \item \textbf{ASSISTments2009: }This variant was sampled from the school year $2009-2010$ with a total number of $525,535$ exercise answering interactions generated by a total of $4,217$ students, a total of $26,688$ questions and a total of $123$ KCs. Nonetheless, duplicates were reported to exist in the collected interactions~\cite{Going16}. Another limitation is that only two-thirds of the questions in this dataset were linked to their relevant KCs and marked with 'NA` in the KCs field. This limitation could burden the use of graph representation learning methods capturing relationships among questions and KCs in the dataset.  
    \item \textbf{ASSISTments2012:} 
    This dataset was collected in the school year $2012-2013$, representing the most significant variant in data volume. It has a total number of $6,123,270$ recorded interactions, a total number of $179,999$ questions, a total number of $46,674$ students, and a total number of $265$ KCs. Despite this large volume of recorded data samples, only $30\%$ of questions were linked to their relevant KCs, limiting the benefit from the large volume of collected data.
    \item \textbf{ASSISTments2015: } This dataset records a total of $708,631$ interactions from the school year $2015-2016$, which was resulted by a total of $19,917$ students answering question from a set of $100$ distinct questions and $100$ of total KCs. Unlike the other variants of the ASSISTments group, there was no metadata on KCs, such as KC description or title provided. Also, the linkage between questions and their relevant KCs was missing. 
\end{itemize}

\subsection{STATICS2011 Dataset}
This dataset was collected from a mechanical engineering course\footnote{\url{https://pslcdatashop.web.cmu.edu/DatasetInfo?datasetId=507}} at the \emph{Carnegie Mellon University} during the Fall semester in the year $2011$. We note that the dataset is publicly available upon author request~\cite{STATICS2011}. It contains $361,092$ interactions answered by $335$ students solving questions from a set of $1,224$ unique questions. The total number of KCs in this dataset is $85$. Despite a large number of recorded interactions, only half of this number includes the answer status (i.e., correct or wrong answer). To overcome this limitation, relevant KT literature that utilized the dataset applied a data preprocessing step to filter out data samples without an answer status~\cite{zhang2017dynamic}. 

\subsection{Junyi Academy Dataset}
This dataset was collected by the \emph{Junyi Academy}\footnote{\url{https://www.junyiacademy.org/}}, which is an online education platform in Taiwan. The data was recorded from high school level Math exercise practice between $2010$ and $2015$ years. There are $25,925,992$ interactions, $247,606$ students, $722$ distinct questions, and $41$ KCs. It is to be noted that a preprocessing step is required to filter out interactions without an answer label, and performing this step results in a reduction of the total number of interactions to $21,571,469$. In addition, the linkage between questions and their relevant KCs is not complete for all the distinct questions in the dataset. 

The Junyi academy released an update for this dataset that is available on Kaggle platform\footnote{\url{https://www.kaggle.com/junyiacademy/learning-activity-public-dataset-by-junyi-academy/tasks}} with data recorded from the year $2018$ to the year $2019$. There are $11,468,379$ interactions, $25,649$ students, and $1,701$ distinct questions. This update constrained interactions to include students who attempted each question only once.

\subsection{Simulated-5 (Synthetic) Dataset}
This is a synthetic dataset\footnote{\url{https://github.com/chrispiech/DeepKnowledgeTracing/tree/master/data/synthetic}} that was not recorded from an actual student practicing but from simulation assuming a set of five KCs. It was proposed to evaluate one of the initial deep KT models named Deep Knowledge Tracing (DKT) proposed by Piech et al.~\cite{piech2015deep}. The data is split into training and testing sets, each containing $50$ distinct questions. Each question was limited to being linked to only one KC and a difficulty level. The authors simulated $4,000$ virtual students to answer each defined question, resulting in $200,000$ interactions. The simulated student agents were following a policy derived from the \emph{Item Response Theory} (IRT)~\cite{foster2017review}. Being synthetic is one of the main limitations of this dataset as it cannot be used to reflect or analyze real student answering behaviors. Moreover, the dataset authors assumed that a question could only have one relevant KC, which is not a realistic assumption.

\subsection{KDDcup Dataset}
This dataset resulted from the KDDcup competition~\footnote{\url{https://pslcdatashop.web.cmu.edu/KDDCup}}~\cite{kdd2010} in $2010$ representing an education data mining challenge. The challenge data was collected from an online education system called ``The Cognitive Tutors'' designed by \emph{Carnegie Learning Inc.}~\footnote{\url{https://www.carnegielearning.com/}} in the United States of America. The data were recorded in the time period $2005-2007$ for students from the age group of $13-14$ years old answering basic school-level Algebra questions. The dataset is divided into two subsets based on time period splits as follows.

\begin{itemize}
    \item \textbf{Algebra 2005-2006:} this subset contains data collected between August $2005$ and June $2006$ with a total of $1,084$ distinct questions answered by $575$ students resulting in $813,661$ interactions. There is a total of $112$ KCs in this subset. Filtering out interactions that do not have linked KCs, the total number of interactions drops to $57.8\%$ after preprocessing.
    
    \item \textbf{Algebra 2006-2007} This subset contains data recorded between $2006$ and $2007$ with a total number of $2,289,726$ interactions generated by $1,840$ students answering $90,831$ distinct questions with a total of  $523$ KCs. Similar to the $2005-2006$ subset, there is a portion of interactions without linked KCs, and after filtering them out, drops the number of interactions to $1,567,072$. We also note that upon inspecting this subset, we found inconsistencies in the timestamp recordings. This might impact the integrity of generated answering sequence and thus reduce the dependability of evaluation results on this subset. 

\end{itemize}

\subsection{EdNet Dataset}
This dataset was released by an online education platform called ''Santa``\footnote{\url{https://www.riiid.co/}} developed by \emph{Riiid} in South Korea. The platform is specialized in preparation for the Test of English for International Communication (TOEIC) exam. One of the distinguishing features of this dataset is the bundle question answering demanding that a set of questions be answered altogether. Another essential feature is recording data from different practicing platforms, such as mobile or web applications, to assess their effect on a student's knowledge state.

The data is divided into four subsets based on the notion of tasks that students are performing. It contains a total number of $95,293,926$ interactions generated by $784,309$ students answering questions from a set of $13,169$ unique questions and a total number of $188$ KCs. The EdNet dataset was designed to provide incremental details about student activities and behaviors. The data recording time period spans over two years. 

\subsection{Limitations of Existing Datasets}

We note that common limitations across the reviewed KT datasets can be summarized as missing the following 1) question tag text in the recorded data, 2) KC-KC relationships, 3) ground truth difficulty of questions, 4) recording answer-related facts such as time taken to answer or the number of choice selection changes, and 5) student's feedback during practice such as their perceived question difficulty or their confidence in their answer. The proposed DBE-KT22 addresses all of these limitations to facilitate the usage of advanced KT techniques and enables a variety of machine learning tasks to be evaluated, such as natural language processing on question and KC textual data, graph modeling on question and KC relational data, curriculum learning based on question difficulty data, or learning cognitive analysis on student's recorded feedback during practice. Table~\ref{tbl:cc} compares existing KT datasets and our proposed dataset on key aspects that highlight the mentioned limitation points.

\begin{table*}
  \caption{Knowledge tracing datasets and their characteristics.}
  \label{tbl:CD}
   \begin{adjustbox}{max width=\textwidth}
  \begin{tabular}{l|c|cccc|c}
    \toprule
     \multirow{2}{*}{\textbf{Dataset}}&\multirow{2}{*}{\textbf{Versions}}&\multicolumn{4}{c|}{Number of }&\multirow{2}{*}{\textbf{Public available}}\\\cline{3-6}&&\textbf{Questions}&\textbf{Students}&\textbf{Interactions}&\textbf{KCs}&\\
     
      \midrule
      ASSISTments&\makecell{2009-2010\\2012-2013\\2014-2015}&\makecell{$26,688$\\$179,999$\\$100$}&\makecell{$4,217$\\$46,674$\\$19,917$}&\makecell{$346,860$\\$6,123,270$\\$708,631$}&\makecell{$123$\\$265$\\$100$}&\makecell{Yes\\Yes\\Yes}\\\hline
 
      STATICS&2011&$ 1,224$&$335$&$361,092$&$85$&No\\\hline
      
      Junyi Academy&2015&$722$&$247,606$&$25,925,992$&$41$&Yes\\\hline
      
      Simulated-5&2015&$50$&$4,000$&$ 200,000$&$5$&Yes\\\hline
      
      KDDcup&\makecell{Algebra 2005-2006\\Algebra 2006-2007}&\makecell{$1,084$\\$90,831$}&\makecell{$575$\\$1,840$}&\makecell{$813,661$\\$2,289,726$}&\makecell{$112$\\$523$}&\makecell{Yes\\Yes}\\\hline
      
      EdNet&\makecell{KT1\\KT2\\KT3\\KT4}&\makecell{$13,169$\\$13,169$\\$13,169$\\$13,169$}&\makecell{$784,309$\\$297,444$\\$297,915$\\$297,915$}&\makecell{$ 95,293,926$\\$56,360,602$\\$89,270,654$\\$ 131,441,538$}&\makecell{$188$\\$188$\\$293$\\$293$}&\makecell{Yes\\Yes\\Yes\\Yes}\\\hline

      DBE-KT22&2022&212&1,361&167,222&98&Yes\\
       \bottomrule
\end{tabular}
\end{adjustbox}
\end{table*}

\begin{table*}
\caption{Comparing essential aspects for KT datasets.}
\label{tbl:cc}
  \begin{adjustbox}{max width=\textwidth}
  \begin{tabular}{l|c|c|c|c|c}

    \toprule
   \multirow{2}{*}{\textbf{{Dataset}}}& {\textbf{Question}} & \multicolumn{2}{c|}{\textbf{Relationships}}   & \textbf{Question} & \textbf{Answer} \\\cline{3-4} &\textbf{Text}&\textbf{Question - KC}&\textbf{KC - KC}&\textbf{Difficulty} & \textbf{Confidence}\\
     \midrule
    ASSISTments&$\times$&\checkmark&$\times$&\checkmark&$\times$\\\hline
    STATICS&$\times$&\checkmark&$\times$&$\times$&$\times$\\\hline
     Junyi Academy&$\times$&\checkmark&$\times$&\checkmark&$\times$\\\hline
    Simulated-5&$\times$&\checkmark&$\times$&$\times$&$\times$\\\hline
     KDDcup&$\times$&\checkmark&$\times$&$\times$&$\times$\\\hline
     EdNet&$\times$&\checkmark&$\times$&$\times$&$\times$\\\hline 
     \textbf{DBE-KT22}&\textbf{\checkmark}&\textbf{\checkmark}&\textbf{\checkmark}&\textbf{\checkmark}&\textbf{\checkmark}\\
  \bottomrule
\end{tabular}
\end{adjustbox}
\end{table*}

\section{The DBE-KT22 Dataset}
\label{sec:anukt22}
\begin{figure}
\centering
\vspace*{0cm}
\includegraphics[width=\textwidth, height=120pt]{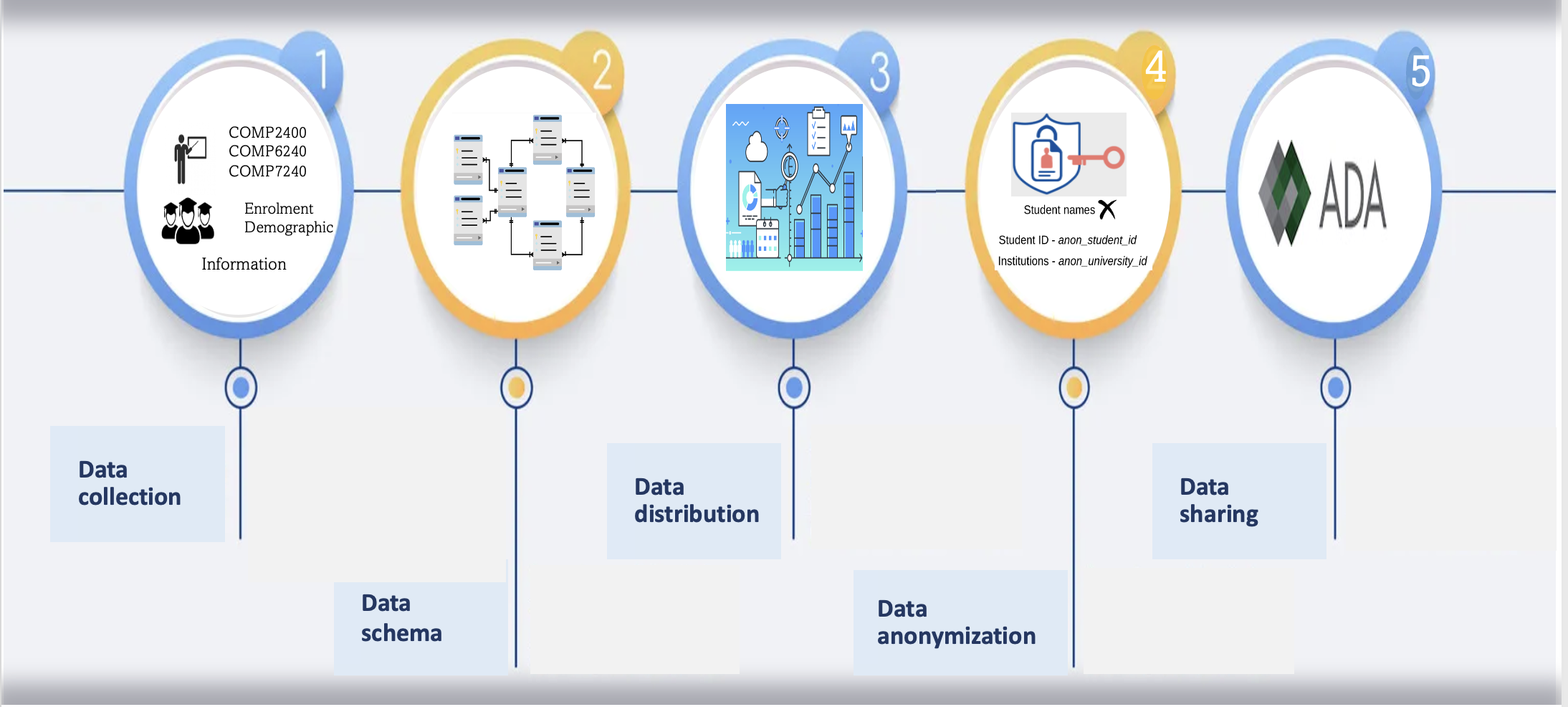}
\caption{The DBE-KT22 dataset production steps.}
\label{fig:mp}
\end{figure}

In this section, we introduce the methodology for producing the DBE-KT22 dataset including the data collection procedure, data relational schema, data distribution, privacy preservation procedure, and data sharing. Additionally, we present the key statistical characteristics of the dataset and visualize them using graphs. Figure~\ref{fig:mp} presents our methodology for producing the DBE-KT22 dataset. In the data collection step, we elaborate on the means to collect the data, the types of collected data, and aspects of the participation environment. The structure of the collected data and their relational dependencies is introduced in the data schema step. The data distribution step presents abstract statistical aspects of the collected data. We clarify the followed procedure to preserve participants' privacy in the data anonymization step. Finally, we illustrate the procedure for sharing our collected data and making it publicly available for interested audiences in the data sharing step.

\subsection{Data Collection}
The DBE-KT22 dataset was collected based on real student exercise practicing in the \emph{Relational Databases} course taught at the Australian National University (ANU) in Australia. Students were mainly at the undergraduate level from different specializations including computer science, engineering, science, business, economics, arts, social sciences, and law and humanities. We note that to the best of our knowledge, our dataset is the first to cover such broad undergraduate student groups across the available KT datasets. The data was collected over a three-year time period from $2018$ to $2021$.

\begin{figure}[t!]
\begin{minipage}{0.48\linewidth}
\centering
\hspace*{0cm}\fbox{\includegraphics[scale=0.18]{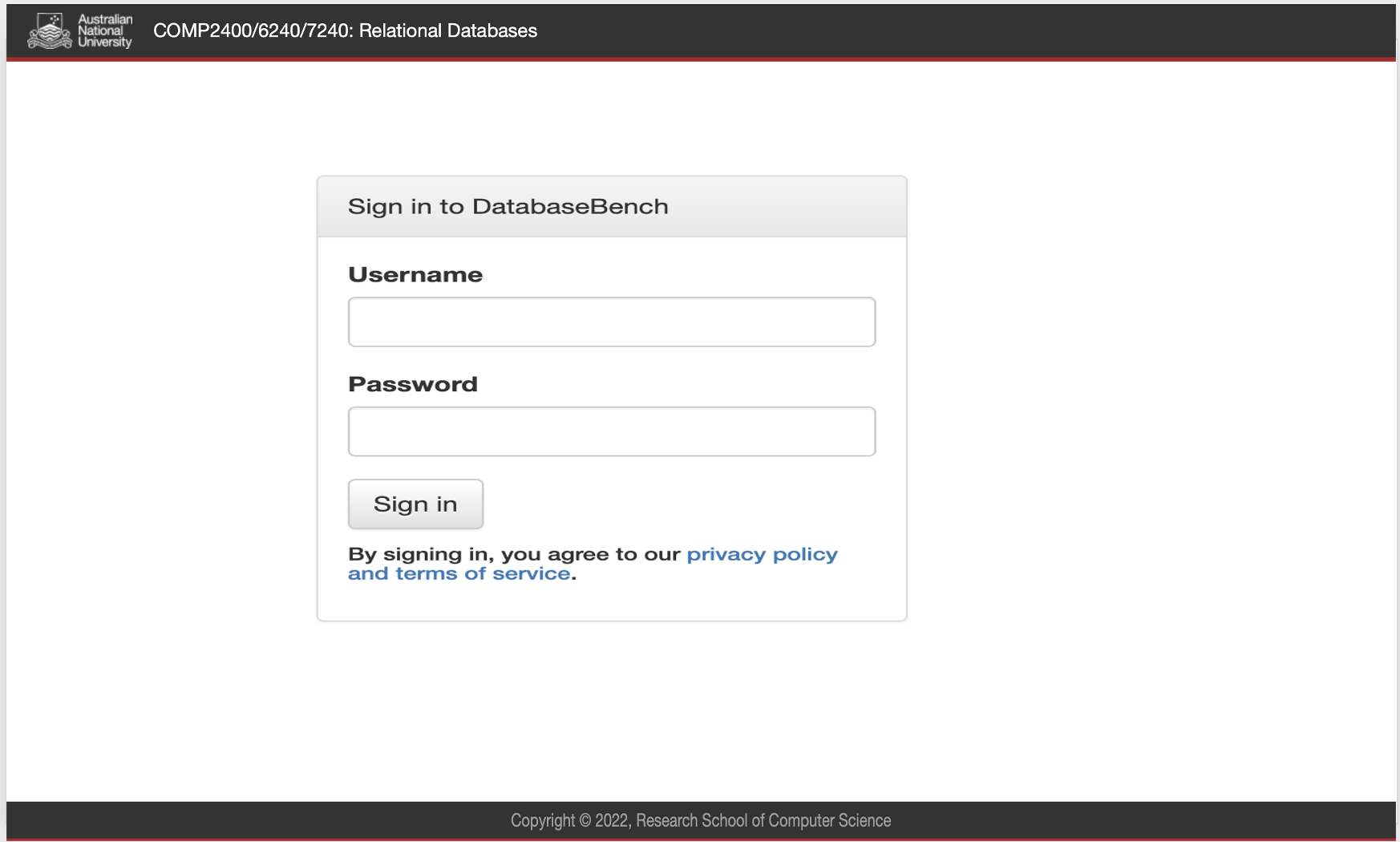}}
\vspace{-0cm}
\begin{center}
(a)  
\end{center}
\end{minipage}
\begin{minipage}{0.48\linewidth}
 \centering
\hspace*{0.1cm}\fbox{\includegraphics[scale=0.18]{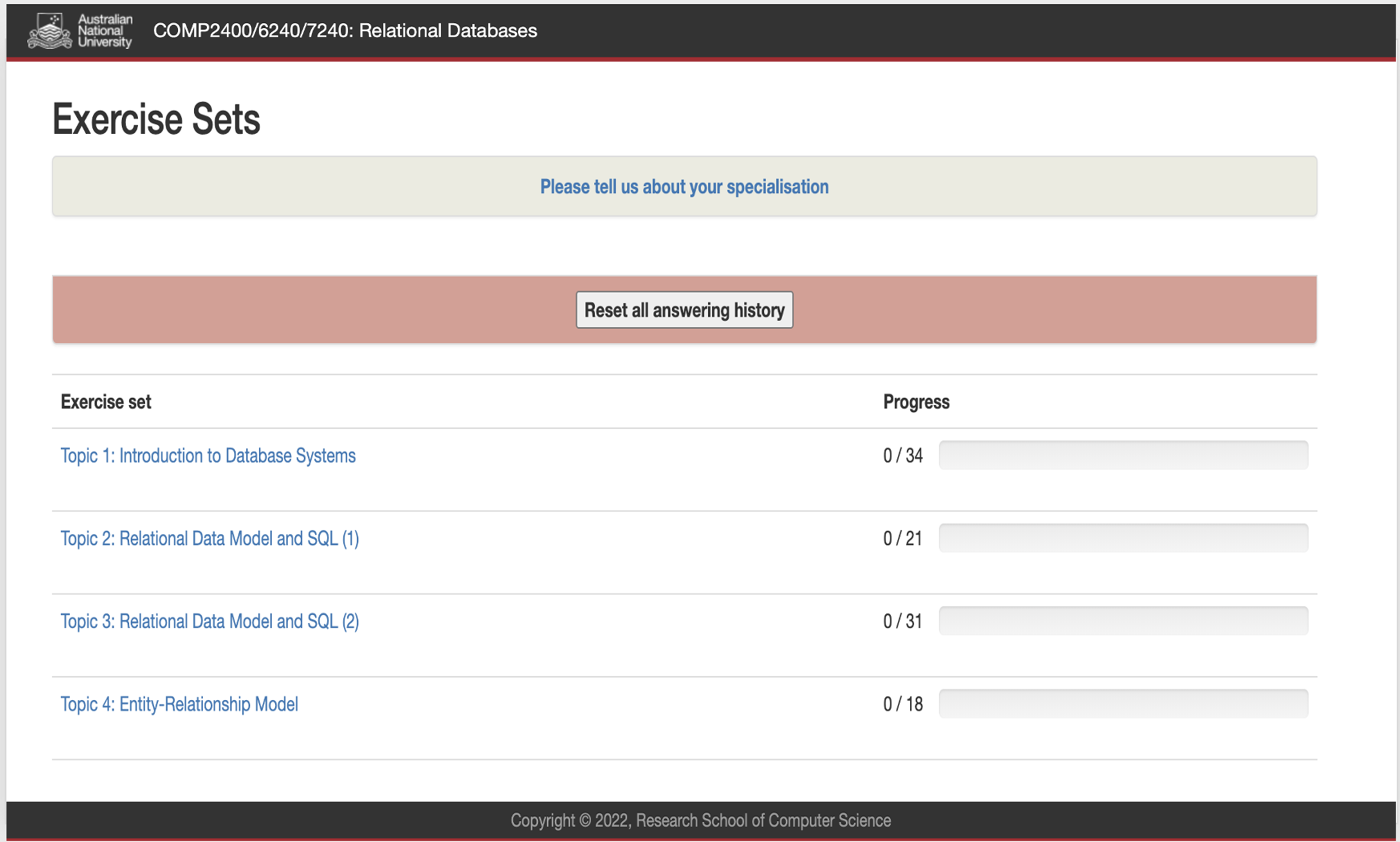}}
\vspace{-0cm}
\begin{center}
(b)  
\end{center}
\end{minipage}
\begin{minipage}{0.48\linewidth}
 \centering
\hspace*{0.1cm}\fbox{\includegraphics[scale=0.18]{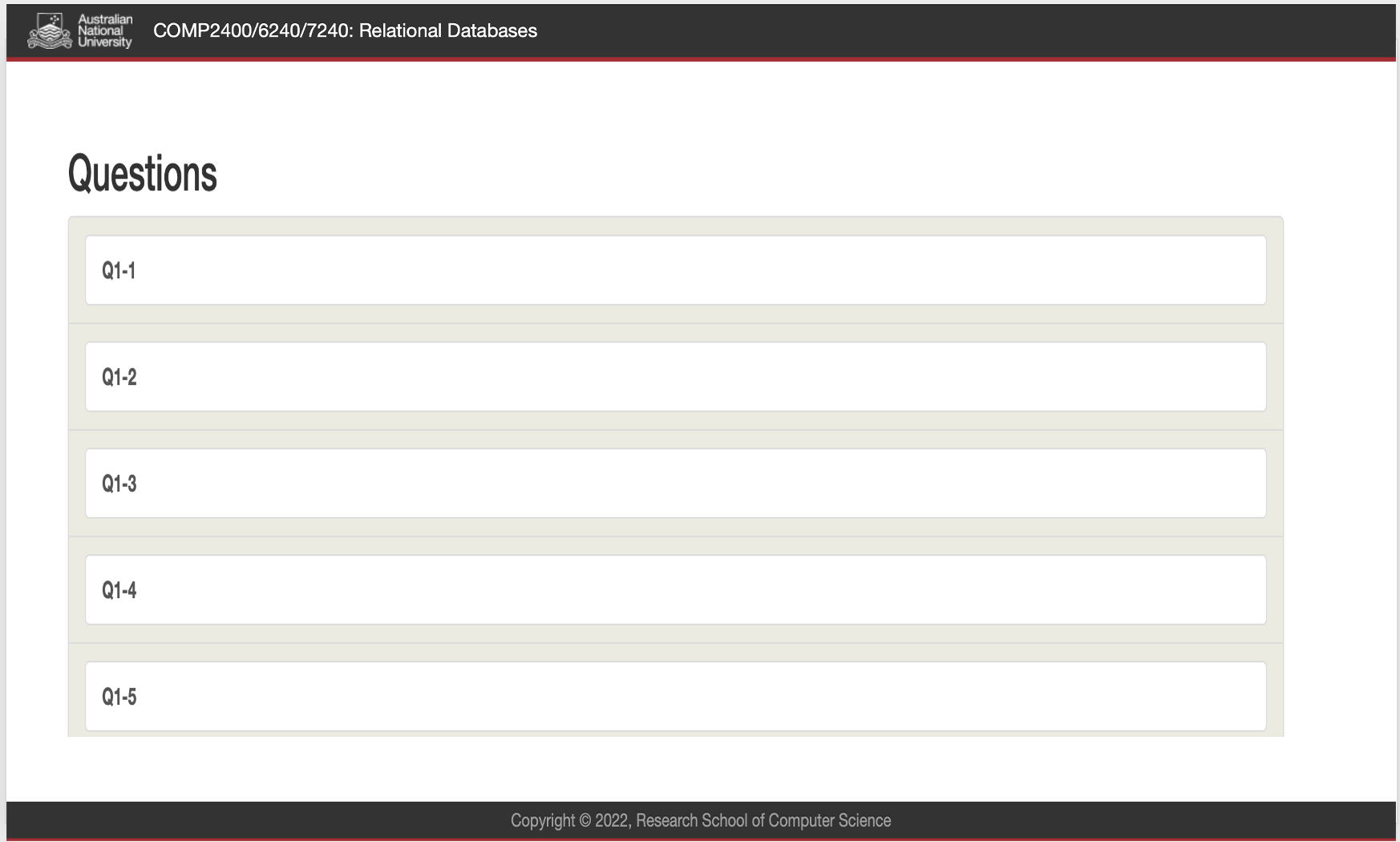}}
\vspace{-0cm}
\begin{center}
(c)  
\end{center}
\end{minipage}
\begin{minipage}{0.48\linewidth}
 \centering
\hspace*{0.1cm}\fbox{\includegraphics[scale=0.18]{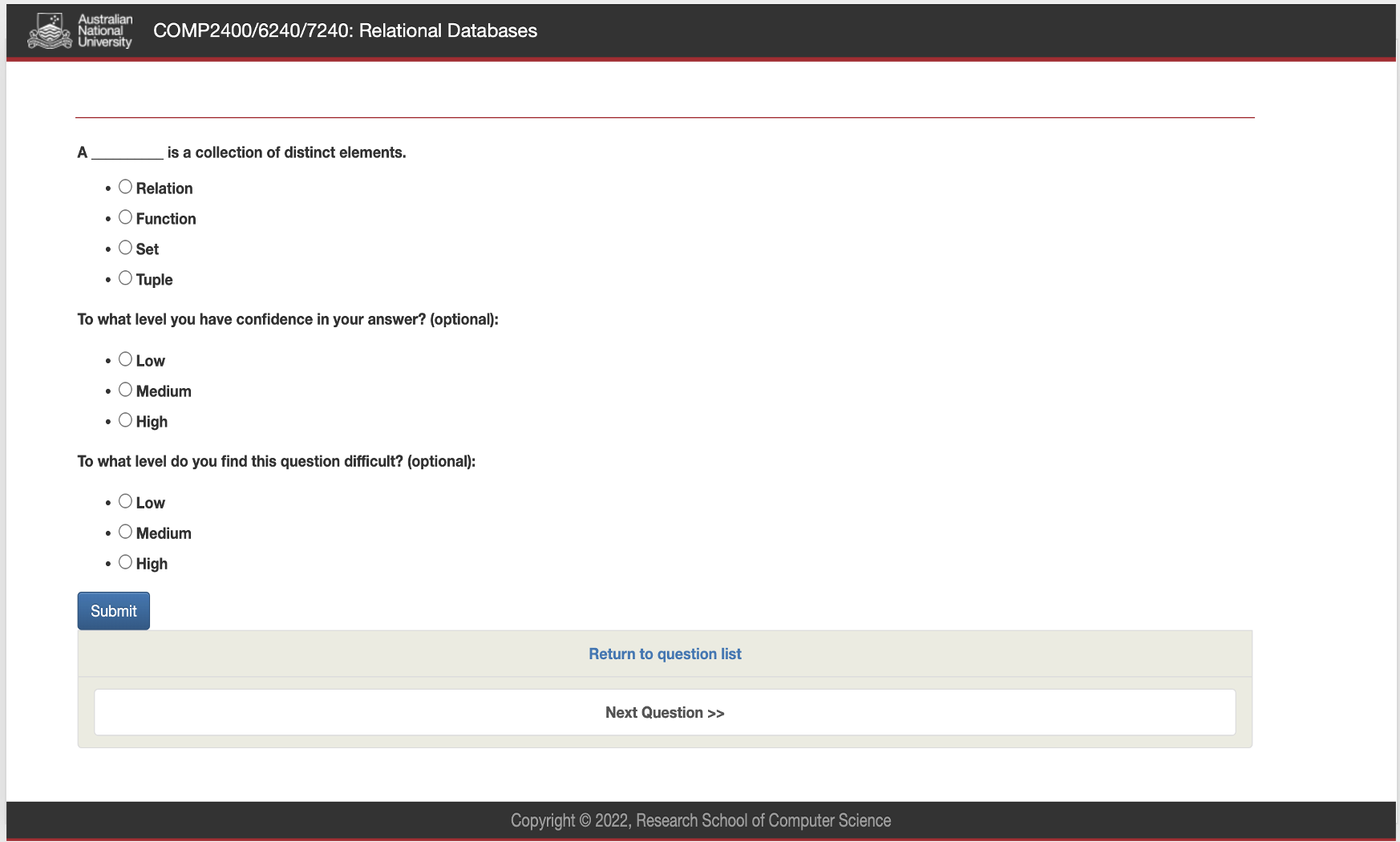}}
\vspace{-0cm}
\begin{center}
(d)  
\end{center}
\end{minipage}
\vspace{-0cm}
\caption{Screenshots for the \emph{CodeBench} web platform for exercise answering. (a) login page. (b) weekly exercise set page. (c) questions list per exercise set. (d) question answering page.}
\label{fig:cobe}
\end{figure}

We utilized an exercise practicing platform developed by the ANU named the \emph{CodeBench} platform~\footnote{\url{https://cs.anu.edu.au/dab/index.html}}, which is a web-based application that can be accessed exclusively by ANU students and staff using their university IDs. The platform enables students to practice exercises in a self-paced manner, where exercises are organized by study weeks. Figure~\ref{fig:cobe} shows screenshots of the main web pages in the platform. Firstly, a student would log in using their university credentials. Then, they are redirected to a home page showing weekly exercise sets for the course; for each set, they will be able to track their completion progress shown through a progress par. Once a student selects a specific weekly exercise set, they see all the exercises within it and can select a given exercise to practice. A student is presented on the exercise practice page with a question title, meta-data such as graphs, tables, or images, and answer choices. We note that all of the questions in the dataset are  multi-choice ones. Besides, a student can show a hint if the instructor provided it for the current question. To record a student's perspective on question difficulty and their trust in the answer before submitting it, we provide two optional questions under each exercise, asking the student for their feedback on difficulty and trust in the selected answer. Moreover, we passively record the number of times a student changed their answer selection and the total time taken to submit the answer for additional features about their confidence in the answer.

\begin{figure}
\centering
\vspace*{0cm}
\includegraphics[width=0.85\textwidth]{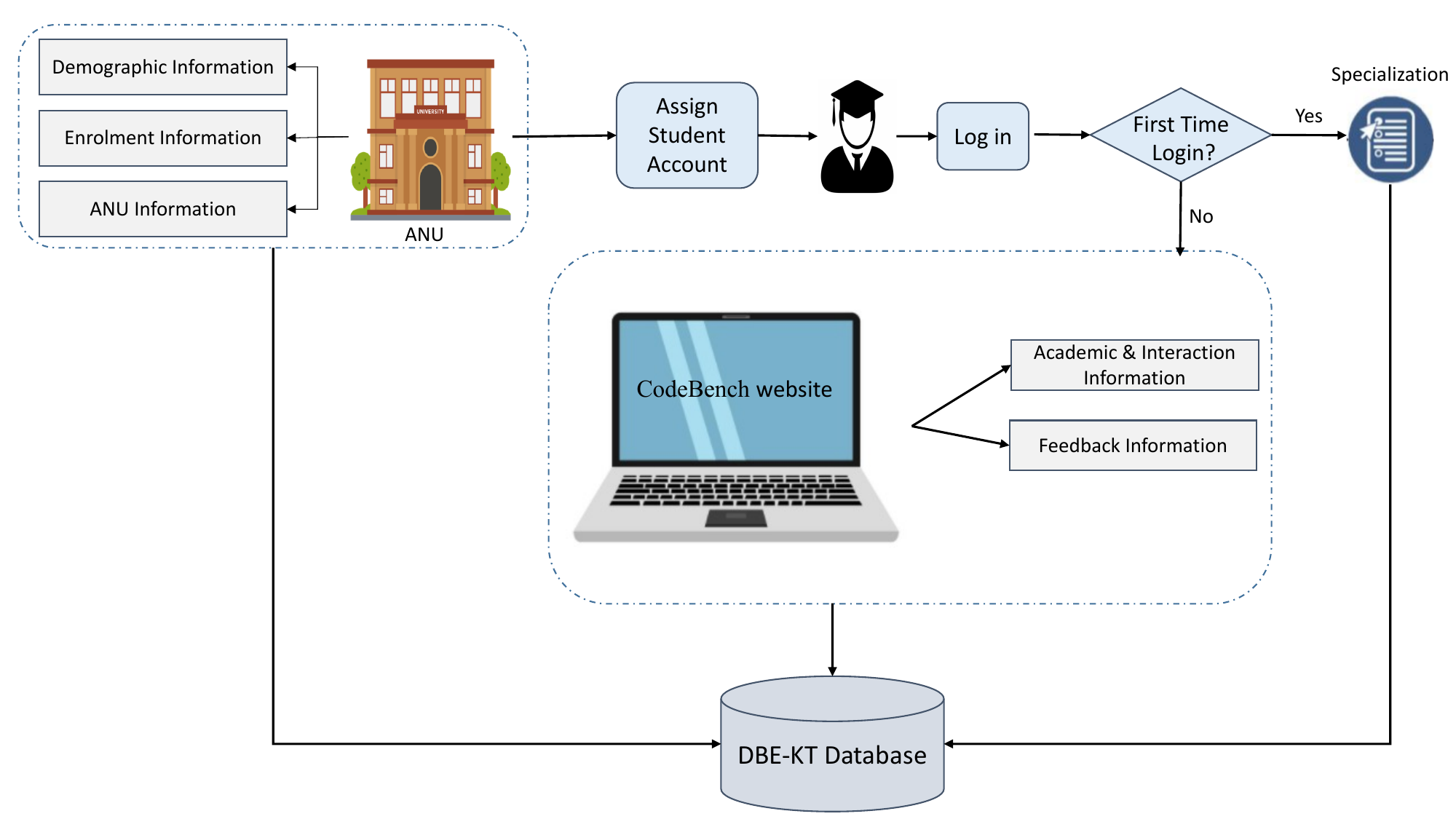}
\caption{DBE-KT22 data collection workflow.}
\label{fig:data_collect_WF}
\end{figure}

Figure~\ref{fig:data_collect_WF} depicts a workflow for collecting the exercise answering data including 1) acquiring university credentials upon enrollment in the course, 2) logging into the \emph{CodeBench} web platform, 3) in case it was the first time to log in, answering a questionnaire about specialization and confirming consent on the course practicing code of conduct, 4) practice exercises, and 5) save practice activity data into the database. We collected practice activity data over three academic years in the period $[2019-2021]$ and ordered the answering sequence for each student chronologically. The schema of the collected data is introduced in the following section.

\subsection{Data Schema}
\begin{figure}
\centering
\vspace*{0cm}
\includegraphics[width=0.75\textwidth]{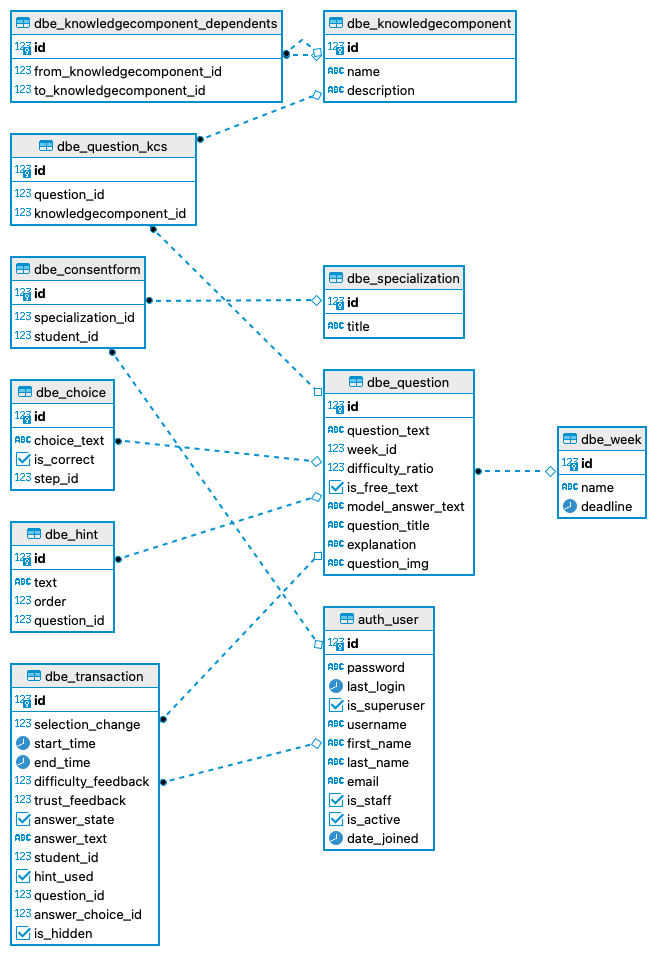}
\caption{Entity-Relationship Diagram for the DBE-KT22 database.}
\label{fig:ERD}
\end{figure}

The schema for the DBE-KT22 dataset follows a relational model that assigns each data aspect to a unique entity (i.e., table) and preserves the relationships across entities using primary-foreign key pairs. Figure~\ref{fig:ERD} shows the entity-relationship diagram (ERD) for the DBE-KT22 database.

We describe each of the entity types in Figure~\ref{fig:ERD} in the following:

\begin{itemize}
    \item \textbf{dbe$\_$question}: this is the main table in this database as it contains all question meta-data.
    \item \textbf{dbe$\_$choice}: this table contains meta-data for choices per each question.
    \item \textbf{dbe$\_$hints}: this table stores hint data for questions.
    \item \textbf{auth$\_$user}: this table contains user login credentials and contact information such as email.
    \item\textbf{dbe$\_$transaction} this table records question answering transaction data including the answer state.
    \item \textbf{dbe$\_$week}: this table contains weekly exercise set data, which are groups of exercises based on a weekly decomposition for the course period.
    \item \textbf{dbe$\_$consentform}: this table stores the data of the student consent including specialization inquiry and confirmation on the course policy.
    \item\textbf{ dbe$\_$specialization}: this is a lookup table for specialization data.
     \item \textbf{dbe$\_$knowledgecomponent}: this table contains meta-data for knowledge components (KCs) in the course.
      \item \textbf{dbe$\_$knowledgecomponents$\_$dependendts}: this table stores relationships between the KCs.
      \item \textbf{dbe$\_$question$\_$kcs}: this table stores relationships between the questions and KCs.
  \end{itemize}
 
 We decomposed the previously described ERD into a set of files to facilitate data sharing and distribution. We followed the Comma Separated Value (CSV) file format for our dataset files for its popularity and the availability of processing software packages. We name each file with the same name as its equivalent table from the ERD. 
 
 Besides, we provide a Python script (i.e., uploaded with the dataset files) to generate training sequences of question answering with relevant meta-data. The script takes three run-time arguments, including one for the desired sequence length to be used in sequencing a student answering history, a padding character for padding sequences shorter than the length argument, and an output file path to save the result. The resultant sequences file named \emph{practice\_sequences.json} is formatted using the JavaScript Object Notation (JSON) format as it is more convenient to represent nested objects such as our sequences samples. The file is structured as an array of JSON objects; each has the following fields:
 
 \begin{itemize}
     \item \textbf{student\_id}: ID of the current student in the sequence.
     \item \textbf{seq\_len}: the length of the current sequence sample. Note that in the case of a sequence with a shorter length than the length argument, this will show the sequence length without counting the padding chars. 
     \item \textbf{question\_ids}: unique ids for questions in the sequence.
     \item \textbf{answers}: answer status for each question in the sequence with $0$ for wrong and $1$ for correct status.
     \item \textbf{gt\_difficulty}: ground truth difficulty level for each question provided by the course instructors. We use $1$ for easy, $2$ for medium, and $3$ for difficult.
     \item \textbf{difficulty\_feedback}: student's feedback on question difficulty before submitting the answer. We use $0$ for not provided, $1$ for easy, $2$ for medium, and $3$ for difficult.
     \item \textbf{answer\_confidence}: student's feedback on their confidence in the selected answer. We use $0$ for not provided, $1$ for low, $2$ for medium, and $3$ for high.
     \item \textbf{hint\_used}: a binary indicator for using hint per each question in the sequence.
     \item \textbf{time\_taken}: time in seconds taken to answer each question in the sequence.
     \item \textbf{num\_ans\_changes}: count for answer selection change per each question in the sequence.
 \end{itemize}

\subsection{Dataset Distribution}

\begin{figure}
\centering
\vspace*{0cm}
\includegraphics[width=0.88\linewidth, height=250pt]{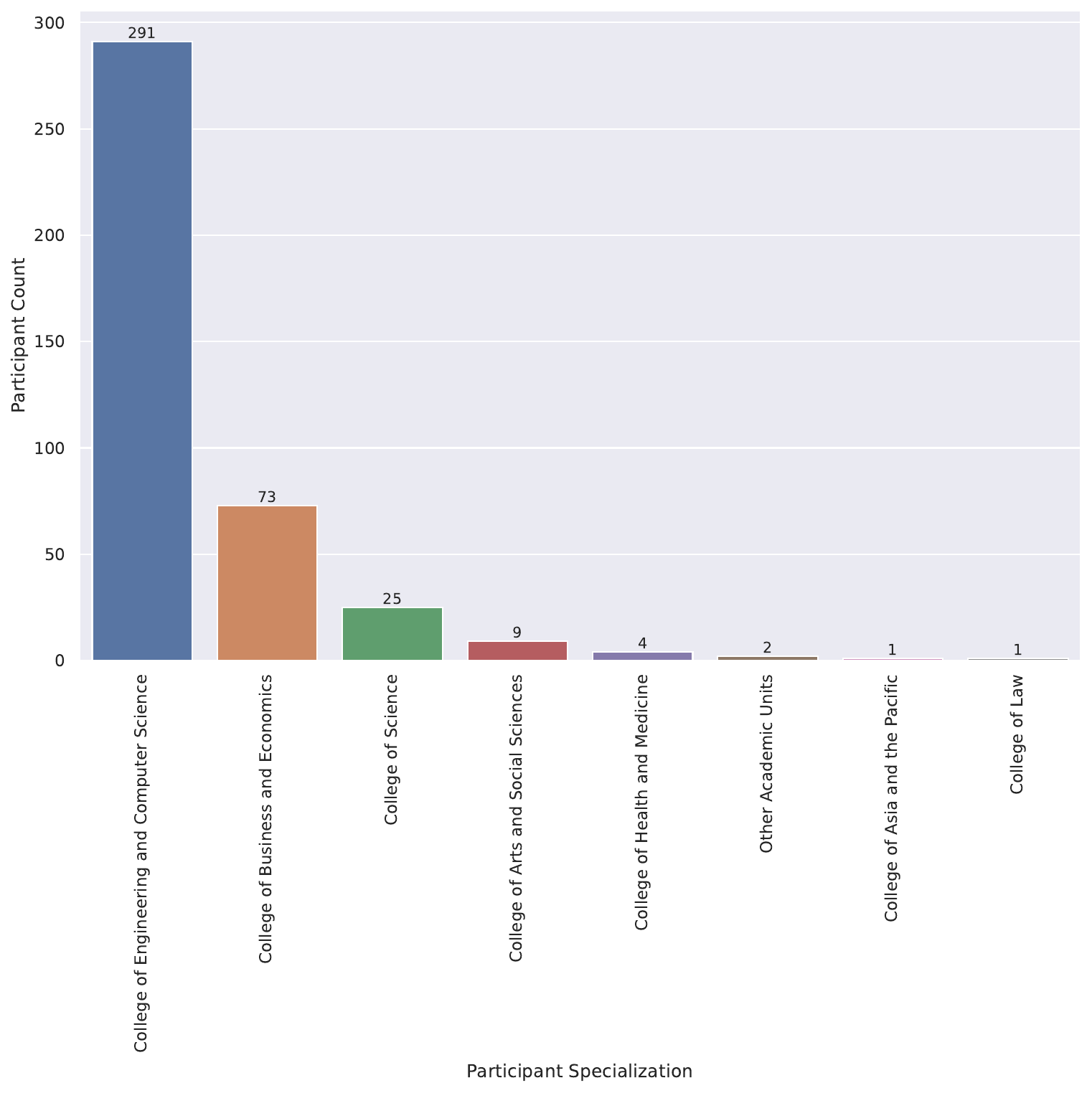}
\caption{The distribution of students' specializations in the DBE-KT22 dataset.}
\label{fig:dist_sd}
\end{figure}

\begin{figure}
\centering
\vspace*{0cm}
\includegraphics[scale=0.5]{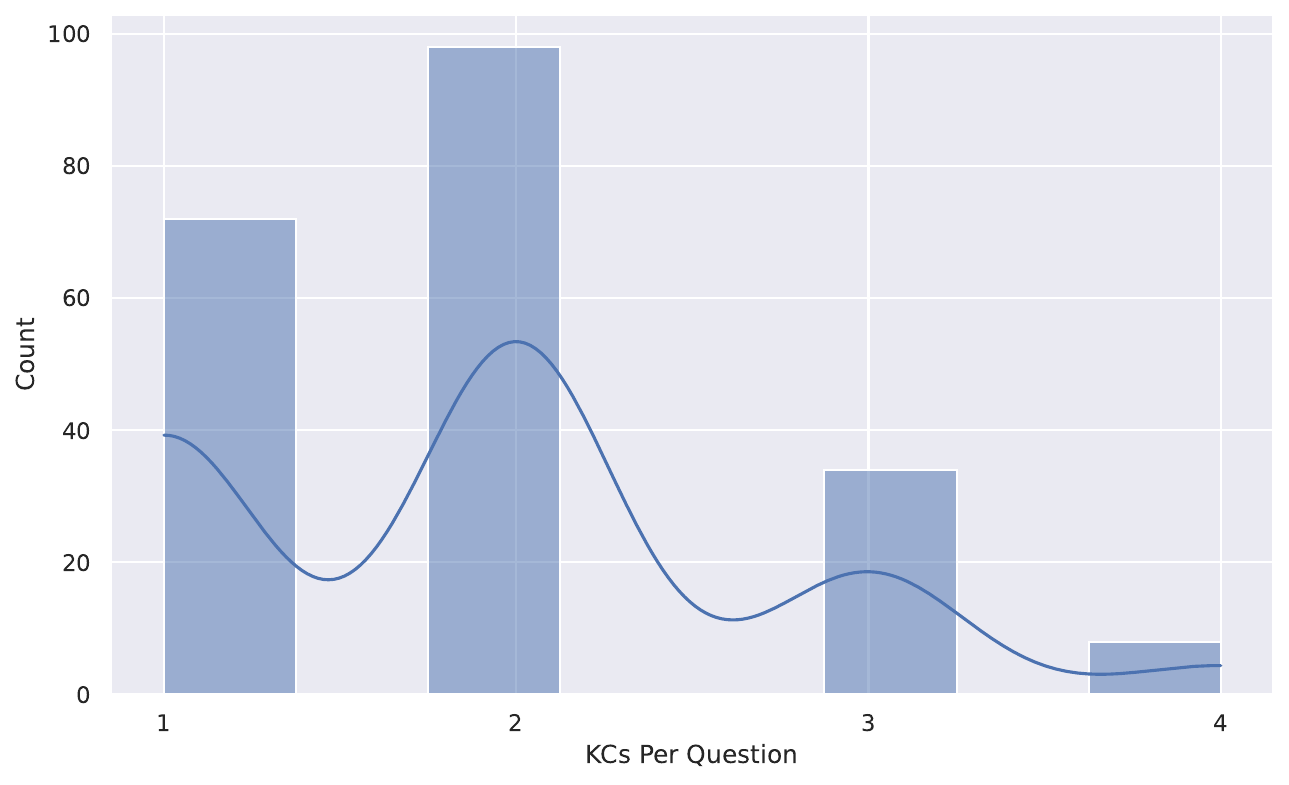}
\caption{The distribution of the KC numbers per question in the DBE-KT22 dataset.}
\label{fig:dist_q_kc}
\end{figure}

\begin{figure}
\centering
\vspace*{0cm}
\includegraphics[scale=0.5]{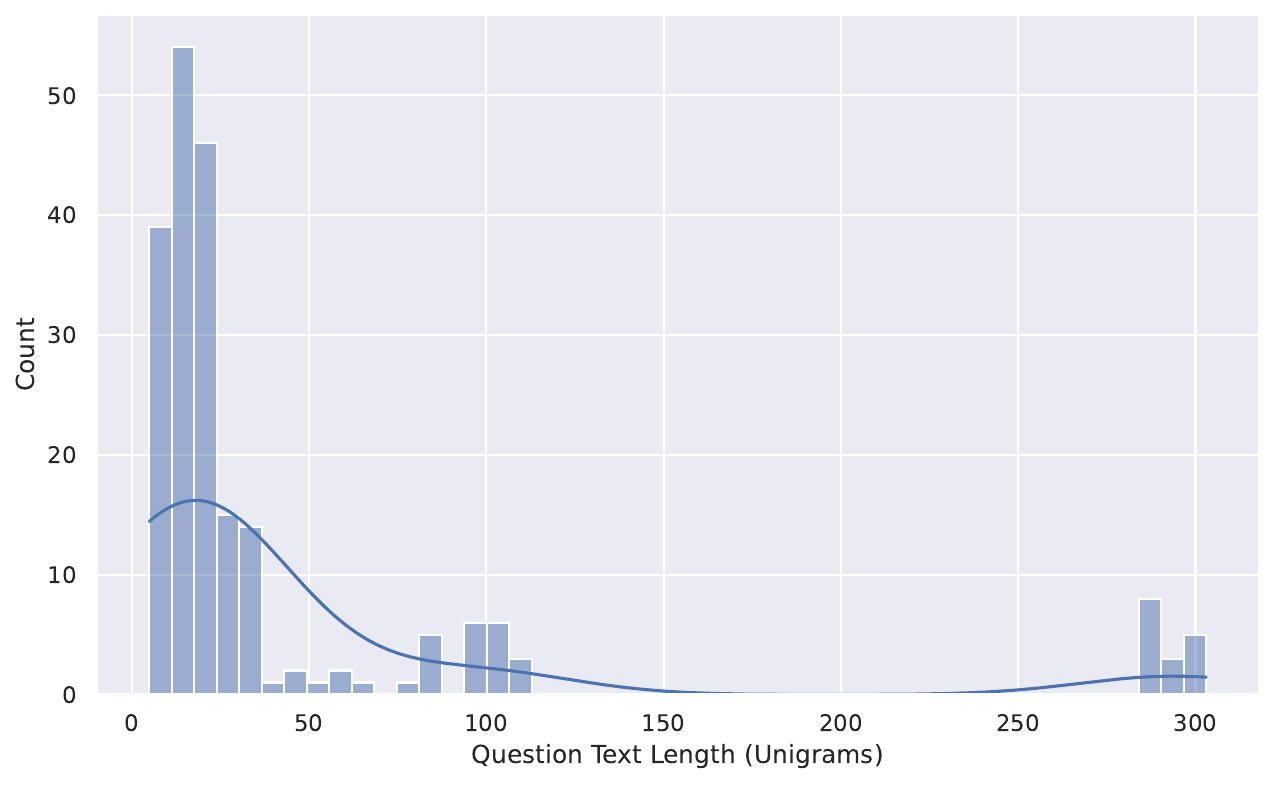}
\caption{The distribution of question text lengths in the DBE-KT22 dataset.}
\label{fig:dist_q_text}
\end{figure}

\begin{figure}[t!]
\begin{minipage}{0.48\linewidth}
\centering
\hspace*{0cm}\includegraphics[scale=0.33]{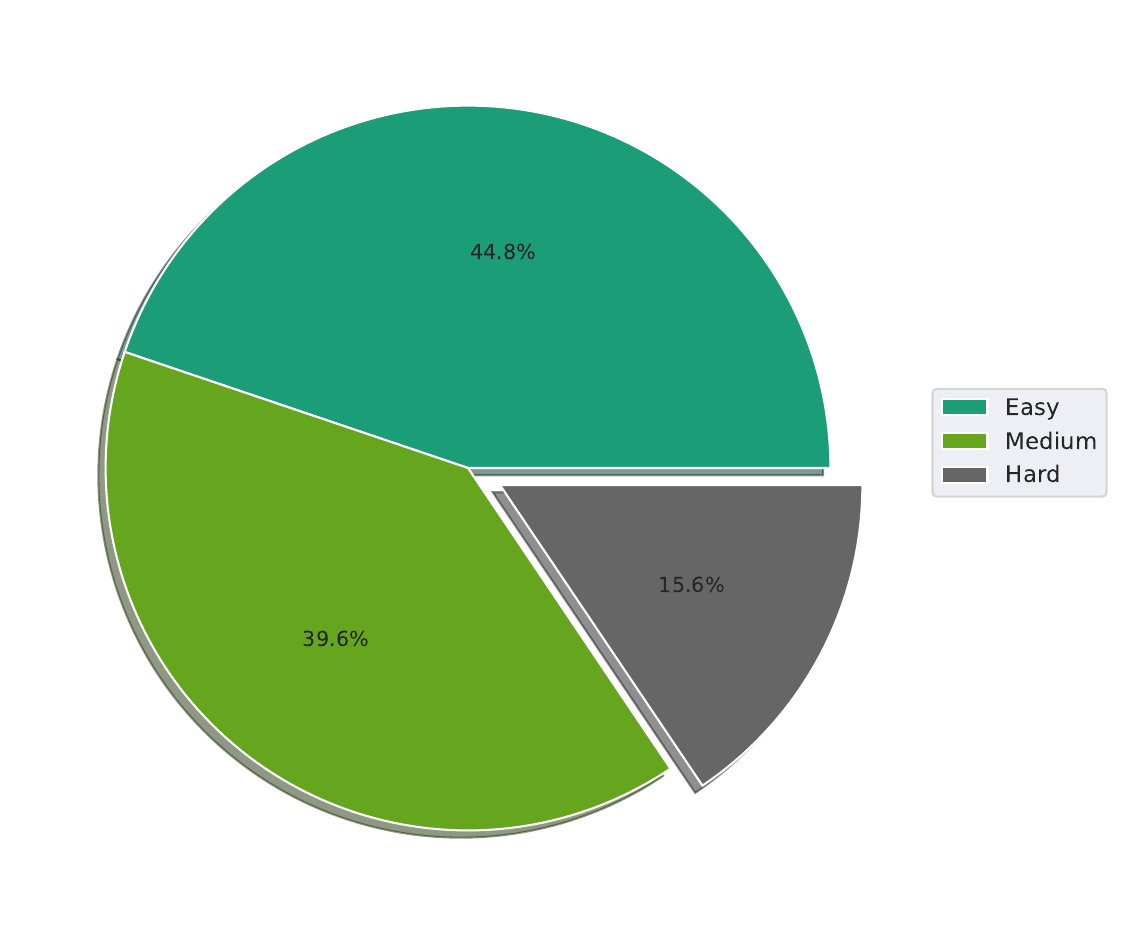}
\vspace{-0cm}
\begin{center}
(a)  
\end{center}
\end{minipage}
\begin{minipage}{0.48\linewidth}
 \centering
\hspace*{0.1cm}\includegraphics[scale=0.33]{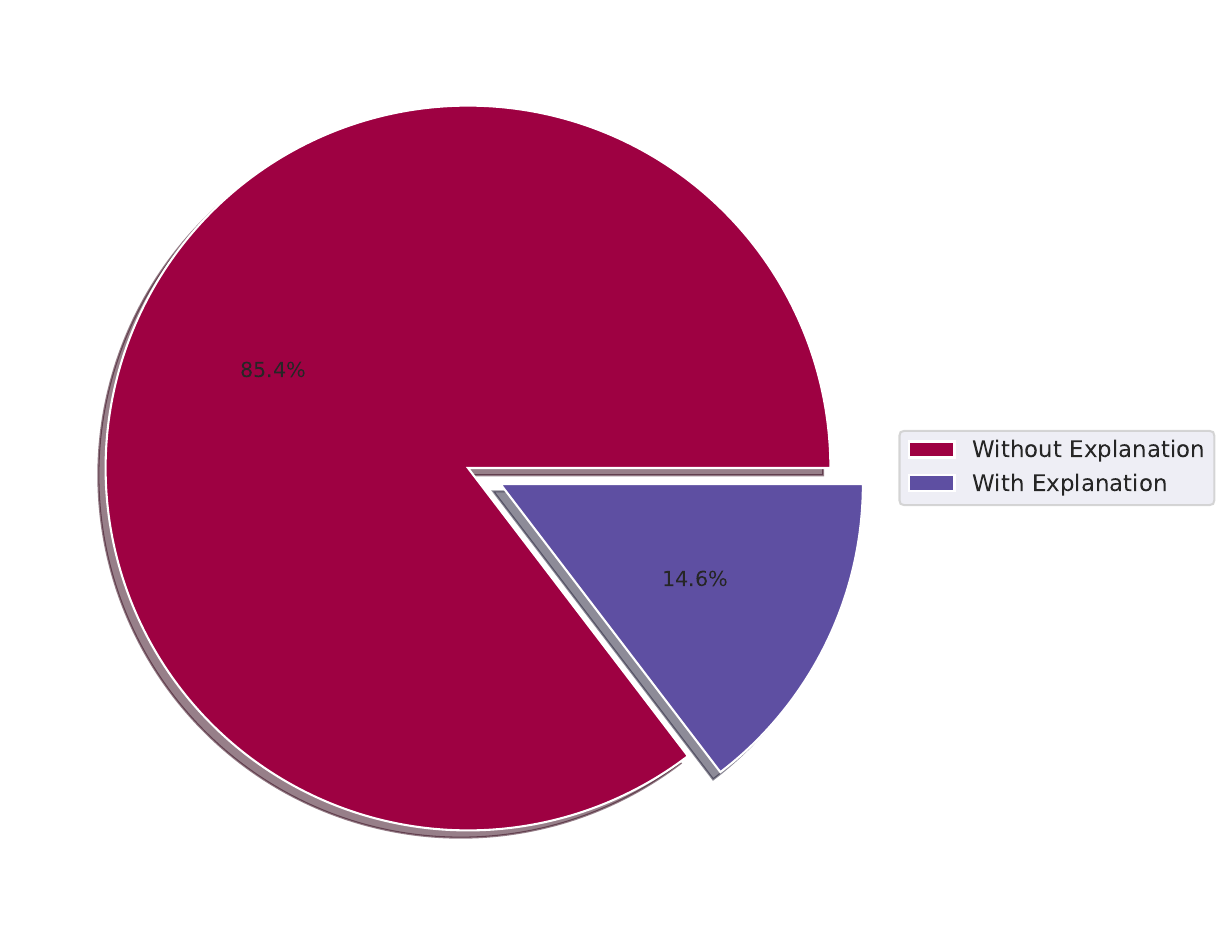}
\vspace{-0cm}
\begin{center}
(b)  
\end{center}
\end{minipage}
\begin{minipage}{\linewidth}
 \centering
\hspace*{0.1cm}\includegraphics[scale=0.33]{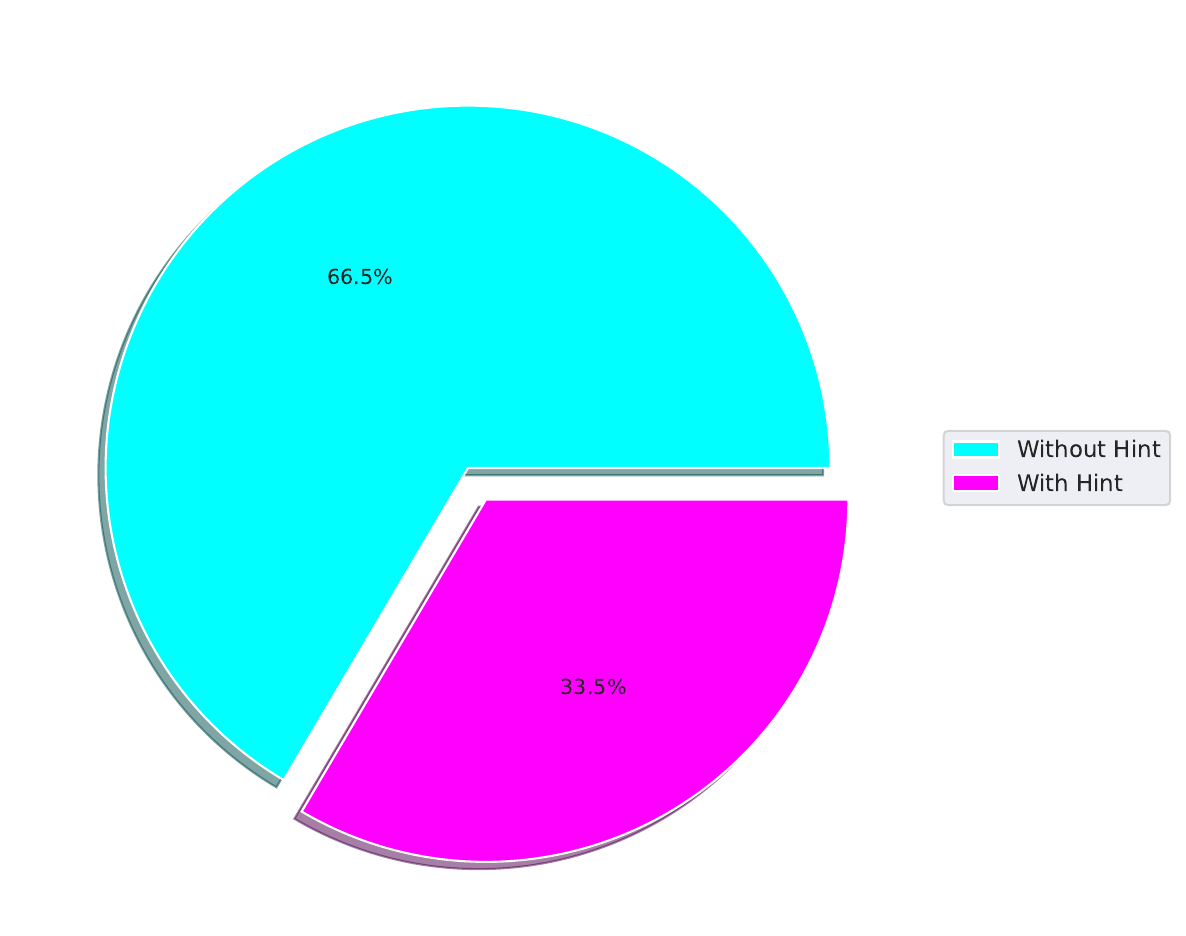}
\vspace{-0cm}
\begin{center}
(c)  
\end{center}
\end{minipage}
\caption{Distribution of question's difficulties, the existence of explanation, and the existence of hint in the DBE-KT22 dataset. (a) difficulty distribution. (b) explanation distribution. (c) hint distribution. }
\label{fig:dist_pies}
\end{figure}

In this section, we use graphs to investigate the data distribution in the DBE-KT22 dataset. We start by showing the distribution of students across specializations. Note that this only includes students who responded to the specialization questionnaire in the consent form. Figure~\ref{fig:dist_sd} shows a bar chart for a visualization of this distribution. The majority of participants come from engineering and computer science specializations. In Figure~\ref{fig:dist_q_kc}, we show a density plot for the distribution of relevant KCs per a given question. It can be observed that most questions in the dataset have $1$ to $2$ relevant KCs. As we include question text in our dataset, it is useful to show the text token length distribution as it is a vital argument for text embedding models. Figure~\ref{fig:dist_q_text} shows a density plot for the token length in the question text. Most questions have a length of less than $50$ tokens, with a minority stretching this number around $300$ tokens. Figure~\ref{fig:dist_pies}.a shows a pie chart for the distribution of question difficulty in the dataset that reflects a reasonable balance between the easy and medium levels and a minority of difficult questions. As per Figure~\ref{fig:dist_pies}.b, the majority of questions (around $85$\%) in the dataset does not have an explanation (i.e., instructor's description of the modal answer), which is aligned with the question difficulty distribution as $84$\% of questions are easy or medium while usually only difficult ones need to have an explanation. The hint (i.e., a small piece of information that could help to approach the answer, yet it couldn't be used as an answer) distribution across questions in Figure~\ref{fig:dist_pies}.c is showing that around $66.5$\% of questions has a hint note to help the student when requested, which covers all the medium and difficult questions in the dataset.

\subsection{Data Anonymization}
While our dataset collects data based on student exercise practice activities, their personal data is usually unrelated to the knowledge tracing context. Thus, we did not aim to collect personally identifying information in our dataset. We only included two fields per each student record, including their ID and specialization. For the IDs, we use an incremental identifier for the dataset records rather than the university-issued ID number to preserve students' privacy. While the specialization field was optional during the data collection process, it can not be used to backtrack students' identities. 

We got our data collection and privacy preservation procedures reviewed and approved by the ANU human research ethics committee~\footnote{\url{human.ethics.officer@anu.edu.au}} under the protocol number 2017/543. 

\subsection{Data Sharing}
 We share the DBE-KT22 dataset files through the \emph{Australian Data Archive} (ADA) platform~\footnote{\url{https://ada.edu.au/}} under an unrestricted access policy. ADA is a platform maintained by the ANU for sharing scientific data with the public and it provides advanced data indexing and search capabilities. Our dataset can be downloaded through its relevant page \footnote{\url{https://dataverse.ada.edu.au/dataset.xhtml?persistentId=doi:10.26193/6DZWOH}} on the ADA that includes general information describing its content, license, and research objectives. The uploaded DBE-KT22 dataset content includes the following files:
\begin{itemize}
    \item \textbf{kc\_relationships.csv}: file contains relationships among KCs. 
    \item \textbf{kcs.csv}: file contains meta-data of KCs. 
    \item \textbf{question\_choices.csv}: file contains choices meta-data per each question. 
    \item \textbf{question\_kc\_relationships.csv}: file contains relationships between questions and KCs.
    \item \textbf{questions.csv}: file contains meta-data for questions, including hint and explanation texts.
    \item \textbf{transaction.csv}: file contains meta-data for question practice attempts.
    \item \textbf{specialization.csv}: file contains meta-data for specializations.
    \item \textbf{student\_specialization.csv}: file contains lookup data linking student ids with their corresponding specializations.
    \item \textbf{sequencer.py}: Python script file for generating training sequences.
    \item \textbf{practice\_sequences.json}: file contains generated training question answering sequences after executing the provided sequencer Python script.
\end{itemize}

\section{Experiments}
\label{sec:experiments}

This section introduces our experimental study investigating critical characteristics of the DBE-KT22 dataset. Mainly, we conduct two experiments, including an exploratory data analysis (EDA) experiment and a question representation learning experiment. The former explores various dataset aspects relevant to the answer prediction task, while the latter evaluates different ways for effective question representation learning.

\subsection{Exploratory Data Analysis (EDA)}
In this experiment, we aim to explore the characteristics of the DBE-KT22 dataset by analyzing the recorded data. We outline six questions investigating the independent variables impacting a student's answer status in the data. Understanding the relationships between these variables and the answer status is vital for any KT usage scenario. The questions are listed as follows.

\begin{itemize}
    \item[-] \textbf{Q1}: How do the question difficulty levels affect the answers of students?
    \item[-] \textbf{Q2}: How does students' answer confidence feedback relate to their answers?
    \item[-] \textbf{Q3}: How well is students' difficulty feedback aligned with the ground truth difficulty of questions?
    \item[-] \textbf{Q4}: How well is students' answer confidence feedback aligned with the ground truth difficulty of questions?
    \item[-] \textbf{Q5}: How dependable is the answering time as an indicator factor for the answers of students? 
    \item[-] \textbf{Q6}: How dependable is the hint usage as an indicator factor for students' answers? 
\end{itemize}

\begin{figure}
\centering
\vspace*{0cm}
\includegraphics[scale=0.6]{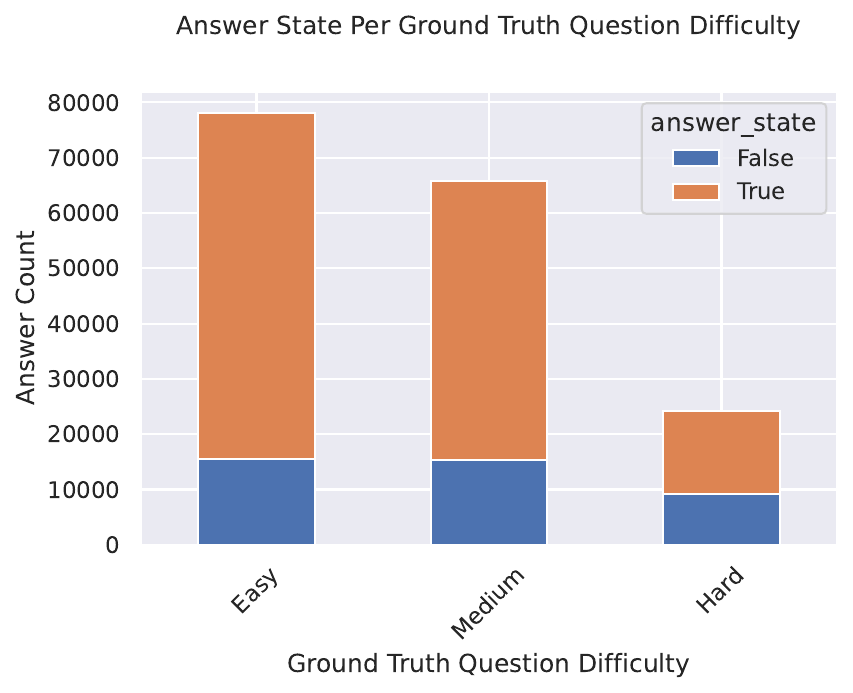}
\caption{A bar chart implying the impact of ground truth question difficulty on the answer status.}
\label{fig:eda_q1}
\end{figure}

\begin{figure}
\centering
\vspace*{0cm}
\includegraphics[scale=0.6]{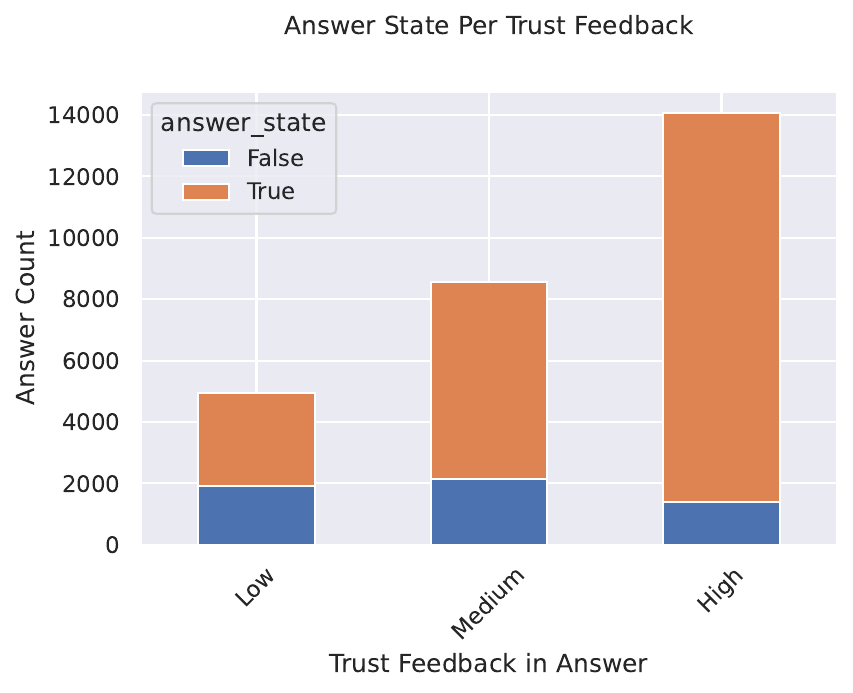}
\caption{A bar chart implying the distribution of answer status over the student's confidence feedback.}
\label{fig:eda_q2}
\end{figure}

\begin{figure}
\centering
\vspace*{0cm}
\includegraphics[scale=0.6]{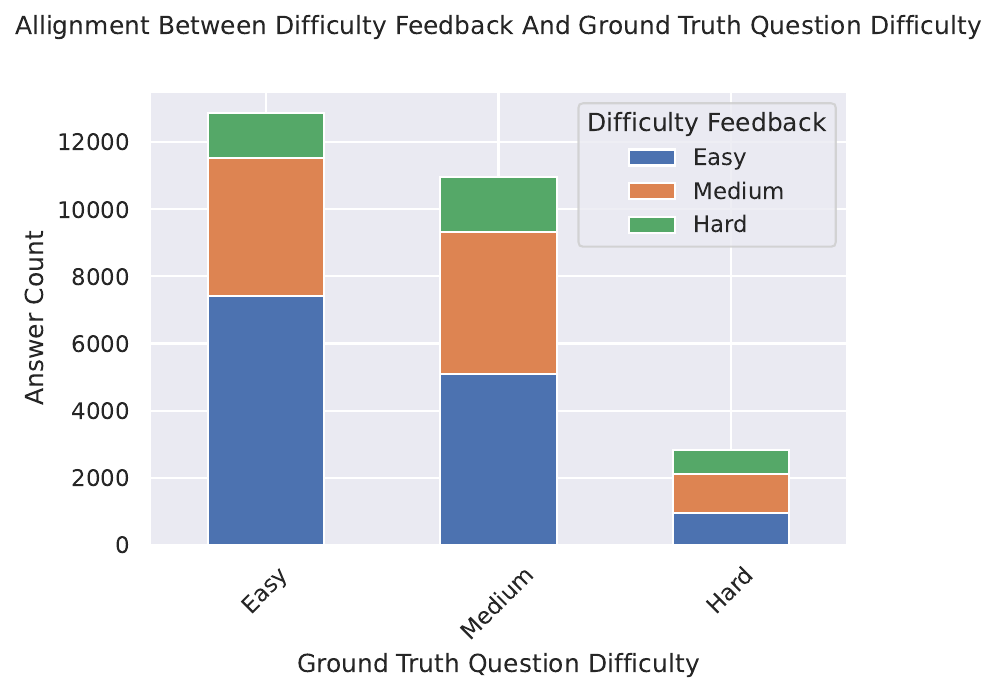}
\caption{A bar chart showing the distribution of student's difficulty feedback over ground truth question difficulty.}
\label{fig:eda_q3}
\end{figure}

\begin{figure}
\centering
\vspace*{0cm}
\includegraphics[scale=0.6]{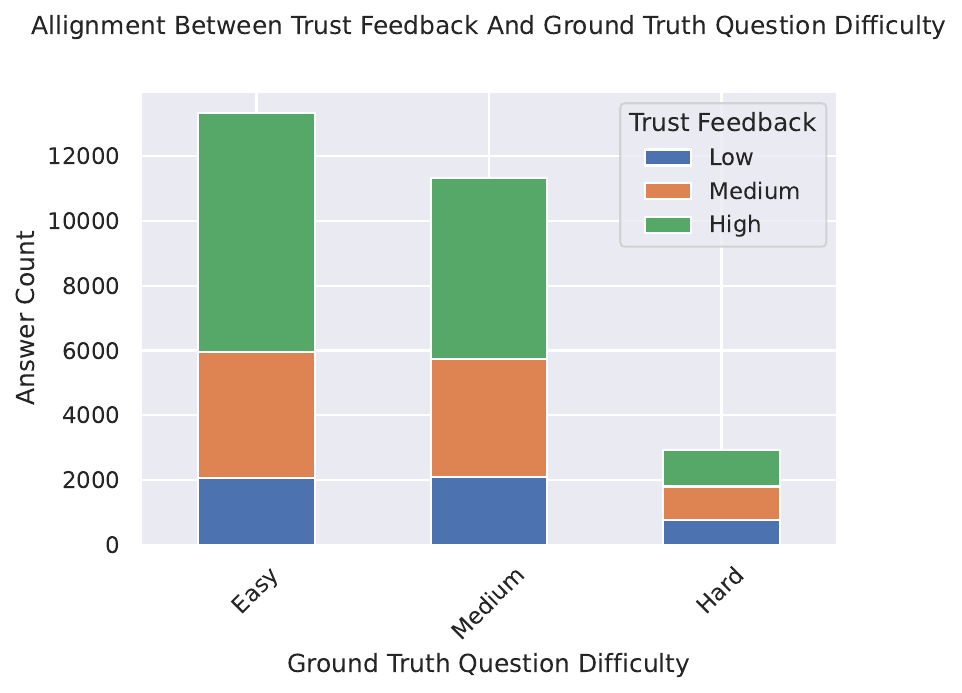}
\caption{A bar chart showing the distribution of student's answer confidence feedback over ground truth question difficulty.}
\label{fig:eda_q4}
\end{figure}

\begin{figure}
\centering
\vspace*{0cm}
\includegraphics[width=\textwidth]{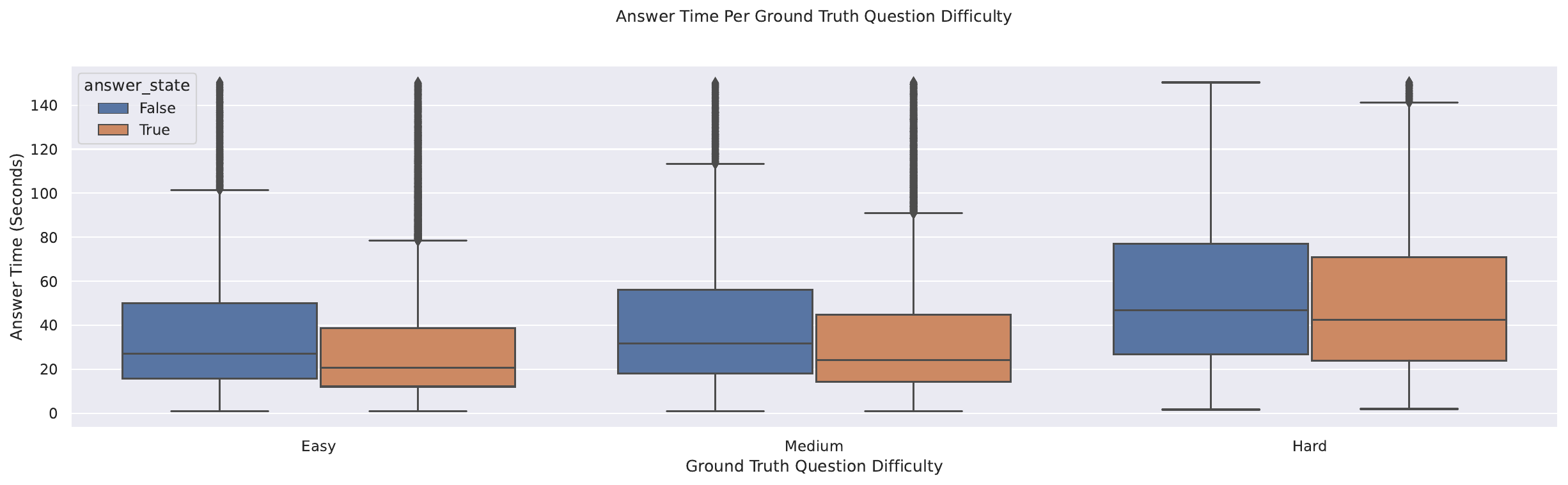}
\caption{A grouped boxplot chart showing the distribution of answering time across answer status and question difficulty levels.}
\label{fig:eda_q5}
\end{figure}

\begin{figure}
\centering
\vspace*{0cm}
\includegraphics[scale=0.5]{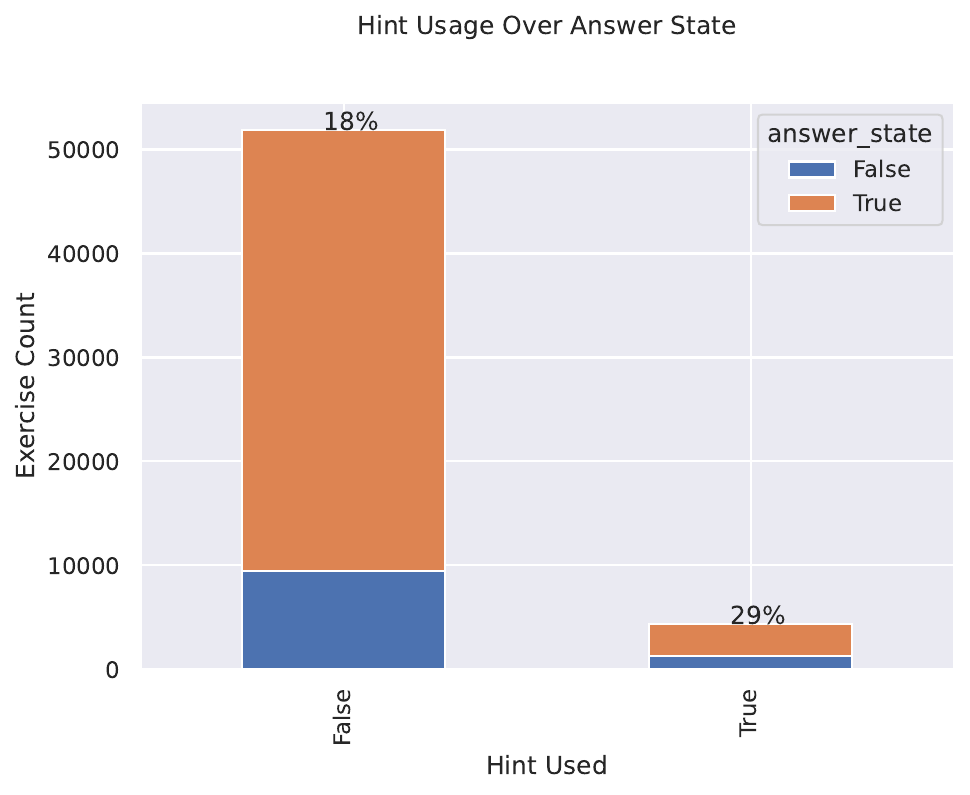}
\caption{A bar plot showing answer status distribution over hint usage. The percentage of wrong answers out of the total answers is shown above each bar.}
\label{fig:eda_q6}
\end{figure}

To answer \textbf{Q1}, Figure~\ref{fig:eda_q1} shows a grouped bar chart for answer status per each ground truth (i.e., provided by the course instructors) question difficulty level. It can be noticed that the distribution between true and false statuses is moving from being imbalanced towards the true status in the easy and medium difficulty levels to a balanced one in the hard level, reflecting the increase of knowledge uncertainty with the increase of question difficulty. To answer \textbf{Q2}, we show the distribution of answer status over student's confidence feedback in Figure~\ref{fig:eda_q2}, which shows an increasing linear pattern in the true answer probability w.r.t the increase in the confidence level. Answering \textbf{Q3}, we investigate the distribution of student's difficulty feedback over the ground truth question difficulty levels. As per Figure~\ref{fig:eda_q3}, one can notice a linear decrease pattern in the number of students reporting an easy question difficulty (blue bars) with the increase of question ground truth difficulty level, which supports the quality of the ground truth labeling done by course instructors.
Similarly, to answer \textbf{Q4}, we inspect the distribution of a student's answer confidence feedback over the ground truth question difficulty. As per Figure~\ref{fig:eda_q4}, we observe that the percentage of high confidence feedback (green bars) linearly decreases with the increase of ground truth question difficulty. For answering \textbf{Q5}, we show in Figure~\ref{fig:eda_q5} the distribution of answering time (i.e., time in seconds taken till submitting the answer) over answer status and ground truth question difficulty levels. We observe that for the easy and medium difficulty levels, the differences between the second quartiles (median) and third quartiles of answering time are significant, with more time taken for false (i.e., wrong) answer status, while these differences are less significant in the hard question difficulty level for the increased uncertainty. Finally, to answer \textbf{Q6}, we show the answer status distribution over hint usage binary indicator in Figure~\ref{fig:eda_q6}. We calculated each bar's percentage of wrong answers (i.e., false status). We observe that using the hint is an effective indicator of a gap in a student's knowledge and would probably imply a wrong answer status, as the percentage of wrong answers is higher when using the hint compared to not using it.

\subsection{Question Representation Learning}
In this experiment, we evaluate different ways of learning a question representation using question text data. We utilize the uncased-base pre-trained BERT~\cite{BERT_19} language model using \emph{Transformers} NLP models hub~\footnote{\url{https://huggingface.co/bert-base-uncased}}. We define the following research questions to guide our evaluation of this experiment:

\begin{itemize}
    \item[-] \textbf{Q1}: What are the different ways to distill question embedding from a pre-trained BERT language model? How effective is each way in clustering relevant questions?
    \item[-] \textbf{Q2}: How to fine-tune pre-trained embedding representation of question text? What is the effect of fine-tuning dataset size on the quality of embedding representation?
\end{itemize}

To answer \textbf{Q1}, we need a reference ground truth embedding representation for questions to evaluate different ways of distilling pre-trained text embeddings. We use the ground truth data on relationships between questions and KCs to design a reference representation. We represent the ground truth on the similarity between a question-question pair by the common KCs between them. Thus, a question in the reference ground truth representation is represented with a binary vector of length $N$ (for the total number of KCs in the dataset), with $1$s in the positions of its relevant KCs and $0$s elsewhere. For distilling question text embedding from the pre-trained BERT model, we evaluate five different ways, including:

\begin{enumerate}
    \item \textbf{CLS token} embedding: represents each question with the CLS token of the final layer in the BERT model.
    \item \textbf{last hidden layer} embedding: represents each question with the CLS token of the last hidden layer in the BERT model.
    \item \textbf{last second hidden layer} embedding: represents each question with the CLS token of the last second hidden layer in the BERT model.
    \item \textbf{last third hidden layer} embedding: represents each question with the CLS token of the last third hidden layer in the BERT model.
    \item \textbf{pooled mean} embedding: represents each question with the reduced mean of the CLS tokens of the last three hidden layers in the BERT model.
    \item \textbf{pooled max} embedding: represents each question with the reduced max of the CLS tokens of the last three hidden layers in the BERT model.
\end{enumerate}

\begin{figure}
\centering
\vspace*{0cm}
\includegraphics[width=\textwidth]{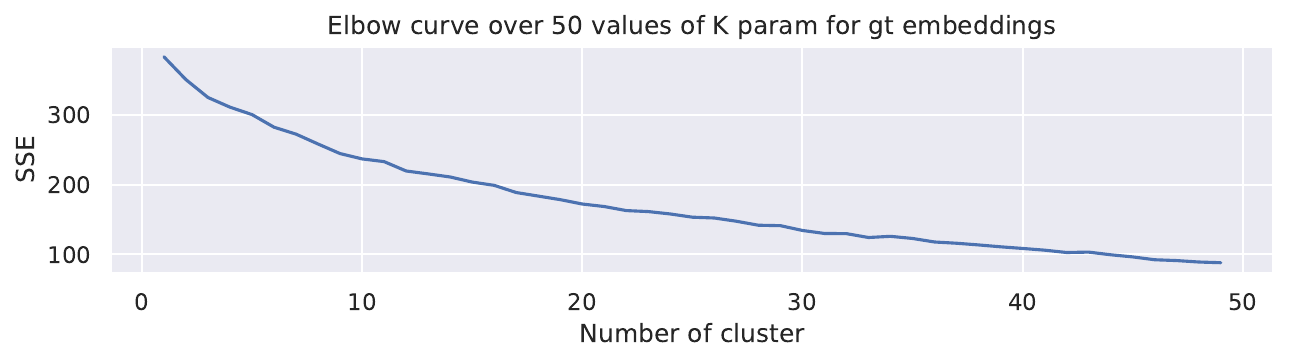}
\caption{The Elbow curve comparing the error sum of squares (SSE) over the number of clusters $K$ using the ground truth question similarity representation.}
\label{fig:embed_elbow}
\end{figure}

\begin{figure}[t!]
\centering
\includegraphics[scale=0.4]{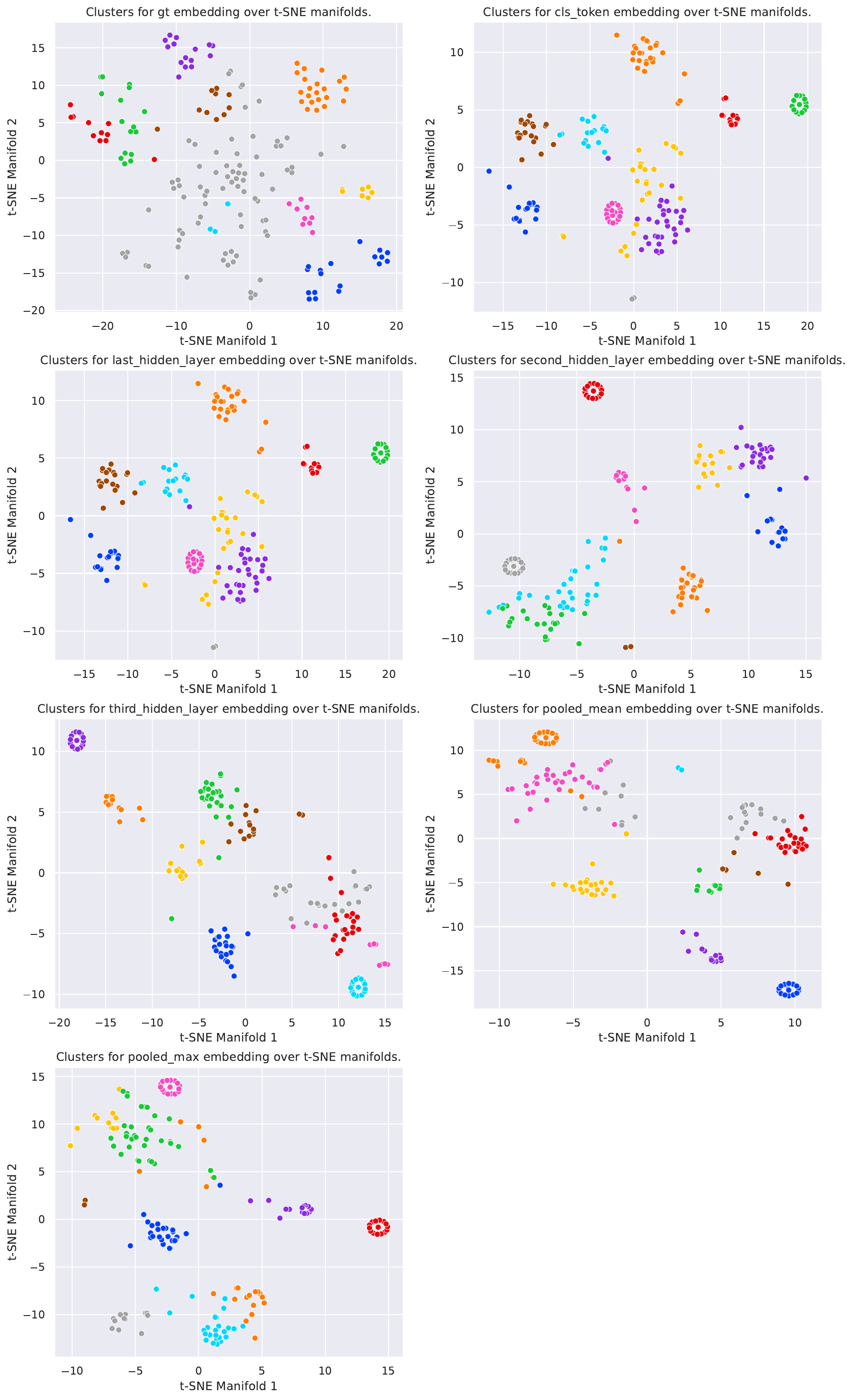}
\caption{A t-SNE two manifold visualization of clustering result for each question embedding method.}
\label{fig:clustering_nontuned}
\end{figure}

We use the K-Means clustering algorithm~\cite{kmean_85} for performing clustering over each embedding method in the evaluation. To decide on the number of clusters $K$, we depend on a data-informed approach using the Elbow method~\cite{elbow_method}. Figure~\ref{fig:embed_elbow} shows a curve for the error sum of squares (SSE) per each $K$ value. SSE is calculated by getting the sum of squared differences between each point and its' cluster mean point. We select $K=10$ as a suitable configuration based on the SSE curve. To visualize the clustering results per each method, we apply t-SNE manifolding~\cite{tsne} on each embedding variant using two manifold components. Figure~\ref{fig:clustering_nontuned} depicts the clustering results for each embedding method. We found that the \textbf{last hidden layer} embedding method outperformed others (see Table~\ref{tbl:clustering_nofinetune}) on the defined clustering performance metrics, including the \emph{Inertia} metric~\cite{kmean_85}, and the \emph{Homogeneity} metric~\cite{homogenity_score}. The \emph{Inertia} metric measures clustering quality by calculating the distance between each data point and its cluster's centroid, squaring this distance, and taking the sum of squares for each cluster. The less the value of this metric, the better the clustering quality. While the \emph{Homogeneity} metric measures the alignment of class labels within a given cluster using a ground truth of class labels. It has a value in the range of $[0,1]$ with $1$ for a perfect alignment and $0$ for a complete dis-alignment. We use the cluster labels generated by the ground truth question representation as class labels to calculate this metric. The higher the value of this metric, the better the clustering quality.

\begin{table}
\begin{center}
\caption{Summary for clustering performance results for ground truth embedding and different methods of question text embedding using Inertia and Homogeneity metrics.}\label{tbl:clustering_nofinetune}%
\begin{tabular}{@{}lcc@{}}
\toprule
Embedding Method & Inertia &  Homogeneity \\
\midrule
Ground Truth Embedding & $203.8$& $1.0$ \\
CLS token embedding & $1923.2$& $0.63$ \\
last hidden layer embedding & $\mathbf{1908.3}$& $\mathbf{0.68}$ \\
last second hidden layer embedding& $1935.0$& $0.59$ \\
last third hidden layer embedding& $2045.6$& $0.58$ \\
pooled mean embedding& $1981.7$& $0.55$ \\
pooled max embedding& $2063.0$& $0.58$ \\
\bottomrule
\end{tabular}
\end{center}
\end{table}

\begin{figure}[t!]
\centering
\includegraphics[scale=0.4]{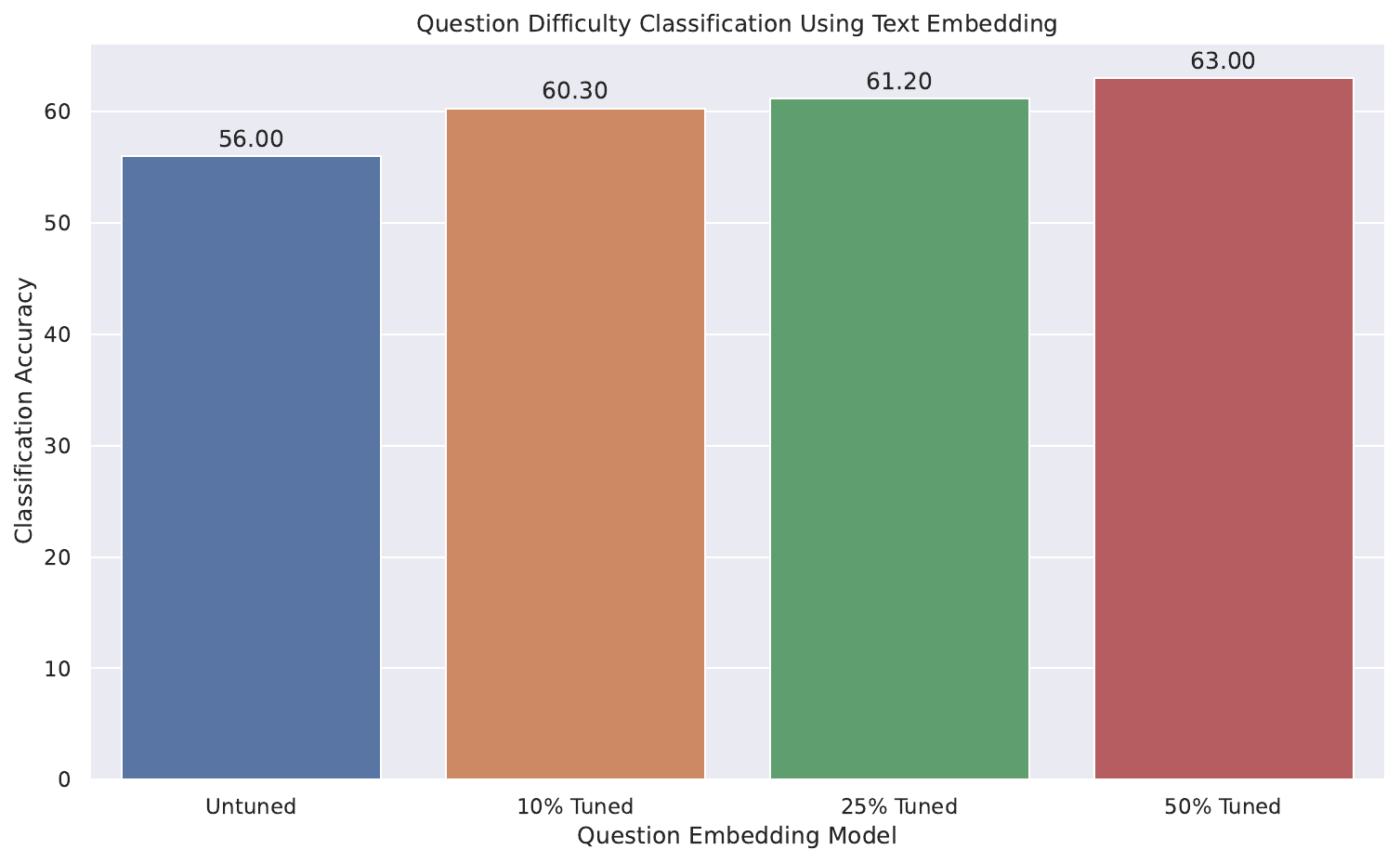}
\caption{A bar chart comparing classification accuracy for different fine-tuned question text embedding variants on classifying question difficulty label task.}
\label{fig:q_diff_predict}
\end{figure}

To answer \textbf{Q2}, we investigate fine-tuning the best question text embedding method from \textbf{Q1} evaluation. We mean by fine-tuning to adapt the embedding parameters of the pre-trained BERT model on our question text data by formulating a supervised learning task to predict the number of shared KCs between a pair of questions using their text embeddings as input. To show the impact of fine-tuning dataset size on performance, we configure three size settings, including $10$\%, $25$\%, and $50$\% of the original training dataset size. This enables us to reflect on the expected enhancement in the learned embedding by introducing more ground truth data about question-KC relationships. We compare the performance of each variant using the same clustering metrics used in \textbf{Q1} evaluation, including \emph{Inertia} and \emph{Homogeneity} metric.
Furthermore, we compare the performance of the fine-tuned embedding variants on classifying question ground truth difficulty label (i.e., $label\in\{1,2,3\}$) from text embedding using classification accuracy metric. As summarized in Table~\ref{tbl:clutering_tuned}, we observe a linear increasing enhancement pattern on the performance of fine-tuned embedding with more ground truth data introduced, yet using $10$\% of the ground truth data size was sufficient to get a significant enhancement with a magnitude of $-37.8$, and $0.03$ for the \emph{Inertia} and \emph{Homogeneity} metrics respectively comparing to the best performer from \textbf{Q1} evaluation. For question difficulty classification results shown in Figure~\ref{fig:q_diff_predict}, we notice a similar linearly increasing pattern in the classification accuracy with more ground truth data added during the fine-tuning procedure and a significant enhancement over the untuned best performer model from \textbf{Q1} evaluation.

\begin{table}
\begin{center}
\caption{Summary for clustering performance results for fine-tuned variants of question text embedding using Inertia and Homogeneity metrics.}
\label{tbl:clutering_tuned}
\begin{tabular}{@{}lcc@{}}
\toprule
Embedding Method & Inertia &  Homogeneity\\
\midrule
Ground Truth Embedding & $203.8$& $1.0$ \\
$10$\%-tuned Embedding Variant & $1870.5$& $0.71$ \\
$25$\%-tuned Embedding Variant & $1845.2$& $0.75$ \\
$50$\%-tuned Embedding Variant & $1830.6$& $0.81$ \\
\bottomrule
\end{tabular}
\end{center}
\end{table}

\newpage
\section{Conclusion}
\label{sec:conclusion}

In this work, we proposed a new dataset for knowledge tracing research named DBE-KT22. The dataset was collected from  real-world exercise answering activities in an undergraduate-level course taught at the Australian National University in Australia within the period from $2019$ to $2021$. The collected data covers a wide range of aspects relevant to the knowledge tracing modeling, including questions meta-data with text, KCs meta-data with text, relationships among questions and KCs, relationships among KCs and KCs, student's feedback on observed difficulty and confidence in the answer, and time taken for answering. This can facilitate more machine learning tasks to be formulated based on our dataset such as answer prediction, question text-aware embedding learning, graph representation learning on relationships between questions and KCs, question difficulty prediction, and cognitive analysis based on student's feedback data. We described the workflow of collecting the dataset and thoroughly investigated its characteristics. Moreover, we performed an experimental evaluation of different methods for learning effective text-aware question embedding from pre-trained language models. Finally, we made our dataset publicly available through the Australian Data Archive (ADA) platform with an unrestricted access license for easier utilization by researchers in the field of knowledge tracing.

\clearpage

 \bibliographystyle{model1-num-names}

\bibliography{main}

\end{document}